\documentclass[aps,prc,twocolumn,floatfix,nofootinbib,showpacs,superscriptaddress,tightenlines]{revtex4-1}

\usepackage{float}
\usepackage{slashed} 
\usepackage{amsmath} 
\usepackage{graphics}      
\usepackage{graphicx}      
\usepackage{longtable}     
\usepackage{url}           
\usepackage{bm}            
\usepackage{epsfig}
\usepackage{amsmath}
\usepackage{lipsum}
\usepackage{amssymb}
\usepackage{amsthm}
\usepackage{bbold}
\usepackage{dcolumn}
\usepackage{epsfig}
\usepackage{color}
\usepackage{xspace}
\usepackage{cancel}
\usepackage{tabularx}
\usepackage[utf8]{inputenc}

\newcommand{\bea}{\begin{eqnarray}}
\newcommand{\eea}{\end{eqnarray}}

\begin{document}

\title{Effect of the Quark-Gluon Vertex on Dynamical Chiral Symmetry Breaking}
\author{M. Atif Sultan}
\email[]{atifsultan.chep@pu.edu.pk}
\thanks{}
\affiliation{Centre For High Energy Physics, University of the Punjab,
Lahore (54590), Pakistan.}
\author{Kh\'epani Raya}
\email[]{khepani@nankai.edu.cn}
\thanks{}
\affiliation{School of Physics, Nankai University, Tianjin 300071, China}
\affiliation{Instituto de Ciencias Nucleares, Universidad Nacional Autónoma de México, Apartado Postal 70-543, C.P. 04510, CDMX, México }
\author{Faisal Akram}
\email[]{faisal.chep@pu.edu.pk}
\thanks{}
\affiliation{Centre For High Energy Physics, University of the Punjab,
Lahore (54590), Pakistan.}
\author{Adnan Bashir}
\email[]{adnan.bashir@umich.mx}
\thanks{}
\affiliation{Instituto de Física y Matemáticas, Universidad Michoacana de San Nicolás de Hidalgo, Edificio C-3, Ciudad Universitaria, C.P. 58040, Morelia, Michoacán, México.}
\author{Bilal Masud}
\email[]{bilalmasud.chep@pu.edu.pk}
\thanks{}
\affiliation{Centre For High Energy Physics, University of the Punjab,
Lahore (54590), Pakistan.}

\date{\today}

\begin{abstract}

In this work, we investigate how the details of the quark-gluon interaction vertex affect the quantitative description of chiral symmetry breaking through the gap equation for quarks. We start from two gluon propagator models widely used in literature and constructed in direct connection with our gradually improved understanding of infrared quantum chromodynamics coupled with its exact one-loop limit. The gap equation is then solved by employing a variety of vertex \emph{Ans\"atze}, which have been constructed in order to implement some of the key aspects of quantum chromodynamics, namely, multiplicative renormalizability of the quark propagator, gauge invariance, matching with perturbation theory in the weak coupling regime, independence from unphysical kinematic singularities as well as manifestly correct transformation properties under charge conjugation and parity operations. On general grounds, all truncation schemes exhibit the same qualitative and quantitative pattern of chiral symmetry breaking, ensuring the overall robustness of this approach and its potentially reliable description of the hadron spectrum and properties. 

\end{abstract}

\pacs{12.38.Aw,12.38.-t,12.38.Lg}
\keywords{Schwinger-Dyson equations, Quark-gluon vertex, Dynamical chiral symmetry breaking}

\maketitle

\date{\today}

\section{\label{sec:intro}Introduction}

If quantum chromodynamics (QCD) is the underlying theory of strong interactions, we
expect all hadronic observables to be calculable from the complete knowledge of the corresponding
Green functions. There is an infinite set of integral field theoretic equations which
describe these $n$-point functions in a coupled and highly non-linear manner.
These are the well-known Schwinger-Dyson equations
(SDEs)~\cite{Schwinger:1951ex,Schwinger:1951hq,Dyson:1949ha}. Their structure is
such that any $n$-point function is related to at least one higher order Green function; the two point one-particle irreducible (1PI) Green functions (propagators)
are related to the three point functions (vertices), which in turn are entangled with the
four-point functions (scattering kernels), {\em ad infinitum}.
In a general formalism, not limited to the perturbative domain,
this infinite set must be truncated by introducing
physically reliable model(s) of some suitable set of Green functions before a solution
becomes tractable. The most favorite choice, which lies on the borderline of a daunting
computational complexity while still maintaining predictable exploration of hadronic physics, is to model the 3-point vertices whether they be quark-photon or quark-gluon interactions~\cite{Ball:1980ay,Curtis:1990zs,Bashir:1994az,Bashir:1995qr,Bashir:2011dp,Albino:2018ncl,Alkofer:2008et,Kizilersu:2009kg,Rojas:2013tza,Aguilar:2014lha,Williams:2014iea,Gomez-Rocha:2015qga,Gomez-Rocha:2016cji,Binosi:2016wcx,Bermudez:2017bpx,Aguilar:2018epe}. 
It is natural to demand any truncation of SDEs to resemble the true dynamics of quarks and gluons to the fullest extent possible, while successfully describing the observable degrees of freedom, namely, mesons and baryons. Several reviews describe the tremendous success of the SDE approach to our continually improved understanding of QCD, hadron spectrum and properties, 
see for example~\cite{Fischer:2006ub,Alkofer:2008tt,Bashir:2012fs,Aznauryan:2012ba,Cloet:2013jya,Horn:2016rip,Roberts:2020hiw}.
Ideally, we can impose the following restrictions on the quark-gluon vertex (QGV)
which enters the gap equation directly and also constrains the kernel of the Bethe-Salpeter equation accordingly~\cite{Chang:2009zb,Qin:2011dd,Binosi:2016rxz}:

\begin{itemize}

 \item The QGV must satisfy the Slavnov-Taylor identity (STI)~\cite{Slavnov:1972fg,Taylor:1971ff}.
This implies that the requirement of gauge invariance fixes the longitudinal part of the quark-gluon interaction,~\cite{Aguilar:2018epe}. Its abelian counterpart is 
generally known as the Ball-Chiu vertex,~\cite{Ball:1980ay}. For most practical purposes and abelian-like truncations, one can start from the Ball-Chiu construction as the longitudinal one and push the remaining information in the rich transverse part of it. 

 \item The transverse part is tightly constrained by the requirements of  
the generalized Landau-Khalatnikov-Fradkin transformations 
(LKFT)~\cite{Aslam:2015nia,DeMeerleer:2018txc,DeMeerleer:2019kmh}
and the transverse Takahashi identities
(TTI)~\cite{Takahashi:1985yz,Kondo:1996xn,He:2000we,Qin:2013mta,Albino:2018ncl,Li:2019xwk,Luo:2019ywn}.

 \item It should reduce to its perturbation theory Feynman
expansion. Note that a truncation scheme of the complete set of SDEs, which maintains 
multiplicative renormalizability (MR) of the quark propagator and
gauge invariance at every level of
approximation, is perturbation theory. Therefore, we expect physically
meaningful solutions of the SDEs to agree with perturbative
results in the weak coupling regime~\cite{Davydychev:2000rt,Bashir:1999bd,Bashir:2000rv,Bashir:2007qq,Bashir:2011vg,Bermudez:2017bpx}.

\item It should transform correctly under the discrete symmetries of 
charge, conjugation, parity and time reversal ($C$,$P$ and $T$).

\item It should be free of any kinematic singularities.

\item It should lead to physical observables which are strictly gauge-independent~\cite{Bashir:1994az,Bashir:1995qr,Bashir:2011dp,Albino:2018ncl}.

\end{itemize}

However, the fact remains that any truncation of the SDEs can only be considered
sensible if it is consistently able to reproduce the experimental observations
pertaining to QCD and hadron physics. 
The leading-order symmetry-preserving rainbow-ladder (RL) truncation achieves that goal quite successfully when studying low lying mesons and baryons~\cite{Chang:2011vu,Eichmann:2016yit,Eichmann:2016hgl}. For instance, since the dynamical chiral symmetry breaking (DCSB) pattern of the pseudoscalar meson sector is governed by a close relationship between the gap equation and the Bethe-Salpeter kernel, supplied by the axial vector Ward-Green-Fradkin-Takahashi identity (WGFTI)~\cite{Qin:2014vya}, it is not a surprise that the RL truncation provides an excellent description of these mesons. 

For the gap equation, the usual practice is to employ models for the gluon propagator, constructed by making connections with lattice results, perturbation theory as well as hadron phenomenology, instead of simultaneously solving the corresponding SDEs. A popular choice is the well-known Maris-Tandy (MT) model~\cite{Maris:1999nt}. This model is composed of two terms: an ultraviolet term, fixed from perturbation theory; and, an infrared enhancement term whose strength is typically determined from the chiral quark condensate. Notice that the MT model was put forward before the SDE prediction for the massive gluon
solution~\cite{Aguilar:2004sw} which was later confirmed in 
lattice studies~\cite{Cucchieri:2007md,Bogolubsky:2007ud,Cucchieri:2010xr,Bogolubsky:2009dc}. It supports a finite but infrared enhanced scalar form factor of the gluon propagator, the so called decoupling solution. It is also in agreement with subsequent SDE and functional renormalization group (RG) results~\cite{Aguilar:2008xm,Boucaud:2008ky,Fischer:2008uz,Aguilar:2009nf,Pennington:2011xs,Blum:2014gna,Cyrol:2016tym,Huber:2017txg},
refined Gribov-Zwanziger
formalism~\cite{Dudal:2007cw,Dudal:2008sp,Dudal:2010tf} and
the earlier suggestion of Cornwall~\cite{Cornwall:1981zr}. Even if one includes the effect of dynamical quarks~\cite{Bowman:2007du,Ayala:2012pb,Aguilar:2012rz}, the qualitative behavior of the gluon propagator remains the same and feeds expected physics back into the gap equation~\cite{Bashir:2013zha}. Those facets of the gluon propagator are confirmed in novel combined continuum and lattice studies~\cite{Binosi:2016nme,Cui:2019dwv}. The Qin-Chang (QC) model~\cite{Qin:2011dd} conveniently captures these infrared qualitative features of the gluon propagator, while the connection with one-loop perturbation theory is still maintained just as it was incorporated in the MT model. 

In this article, we employ the effective coupling of both the MT and the QC
models in association with a set of refined \emph{ans\"atze} for the fermion-boson vertex: 
Ball-Chiu (BC) \cite{Ball:1980ay}, Curtis-Pennington (CP) \cite{Curtis:1990zs},
Kizilersu-Pennington (KP) \cite{Kizilersu:2009kg} and Bashir {\em et al}. (BB) \cite{Bashir:2011dp}. For comparison with these refined vertices 
and the sake of completeness, we have included the results based upon the bare vertex as well.

The article is organized as follows: in Section~\ref{sec:models} we
discuss the preliminaries of the gap equation, introducing
the MT and the QC models. In section~\ref{sec:vertices}, we explicitly discuss
all the vertex constructions we employ, highlighting, comparing and contrasting their merits. Section~\ref{sec:gapequation} details the algebraic expressions for the kernels of the gap equation that stem from the choice of each vertex and Section~\ref{sec:results} contains numerical results as well as
a comparative analysis. Finally, in Section~\ref{sec:conclusions}, we summarize our conclusions and discuss the scope and future applications of this work.

\section{Gap equation: preliminaries and the gluon propagator}
\label{sec:models}
\begin{figure}
\includegraphics[width=9cm]{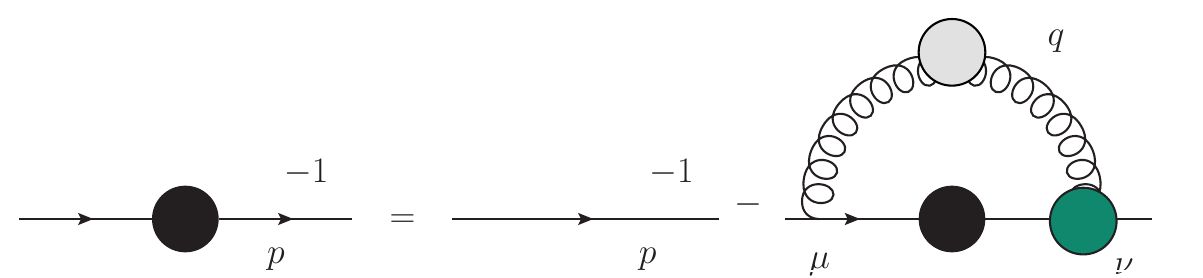}
\caption{SDE of the full quark propagator. The blobs represent fully-dressed propagators and vertices, which obey their own SDEs.}\label{fig:SDE}
\end{figure}

In order to investigate how DCSB is realized, we naturally start from the 
renormalized SDE for the quark propagator. This equation is depicted in Fig.~\ref{fig:SDE} and can be written in the following mathematical form:
\begin{equation}\label{eq:1}
S^{-1}(p)= Z_{2}(i \gamma \cdot p+m_{0}) + \Sigma(p),
\end{equation}
where $\Sigma(p)$ is the quark self energy defined as
\begin{equation}\label{eq:2}
\Sigma(p) = Z_{1} C_{F} \int_k  \;g^2 D_{\mu\nu}(q)\gamma_{\mu}  S(k)    \Gamma_{\nu}(k,p).
\end{equation}
 Here $q=k-p$, $C_{F}=4/3$ and $\int_k \equiv \int^\Lambda \frac{d^{4}k}{(2\pi)^{4}}$. $Z_1$ and $Z_2$ are the renormalization constants for the QGV and the quark propagator, respectively, which depend on the ultraviolet regulator ($\Lambda$) and the renormalization point ($\mu$). This equation, also known as the \emph{gap equation}, involves not only the full quark propagator, $S(p)$, but also the full gluon propagator, $D_{\mu\nu}(q)$, and the fully dressed QGV, $\Gamma_{\nu}(k,P)$. Each of these Green functions also depend on the renormalization point. However, we have not explicitly displayed this dependence for notational convenience. Moreover, they obey their own
 SDEs. 
 This intricate structure yields an infinite tower of coupled equations, which must be systematically truncated in order to extract the encoded physics. Regardless of the truncation scheme, the full quark propagator, representing a Dirac particle, can be defined in terms of two scalar functions, namely the mass function, $M(p^{2})$, and the quark wavefunction renormalization,  $Z(p^{2};\mu^2)$, such that:
\begin{equation}
\label{eq:quarkprop1}
S(p;\mu) = \frac{Z(p^{2};\mu^2)}{i\gamma \cdot p+M(p^{2})},
\end{equation}
in analogy to its bare counterpart
\begin{eqnarray*}
S_{0}(p) = \frac{1}{i\gamma \cdot p+m_{0}}\;,
\end{eqnarray*}
where $m_0$ is the bare mass of the quark. Equivalent useful representations of $S(p;\mu)$ are:
\begin{eqnarray}
\nonumber
S(p;\mu) &=& -  i \gamma \cdot p\;\sigma_v(p^{2};\mu^2) + \sigma_s(p^{2};\mu^2)\;,\\
\label{eq:4}
S^{-1}(p;\mu) &=& i \gamma \cdot p\;A(p^{2};\mu^2) +B(p^{2};\mu^2)\;,
\end{eqnarray}
where the dressing functions involved are interrelated as
\begin{eqnarray*}
\sigma_s(k^2;\mu^2) &=& \frac{B(k^2;\mu^2)}{A^2(k^2;\mu^2)\;k^2+B^2(k^2;\mu^2)} \;,\\
\sigma_v(k^2;\mu^2) &=& \frac{A(k^2;\mu^2)}{A^2(k^2;\mu^2)\;k^2+B^2(k^2;\mu^2)} \;,\\
\end{eqnarray*}
and one can easily identify
\begin{eqnarray*}
M(p^{2}) = \frac{B(p^{2};\mu^2)}{A(p^{2};\mu^2)}\;,\;  Z(p^{2};\mu^2) = \frac{1}{A(p^{2};\mu^2)}\;.
\end{eqnarray*}
Notably, multiplicative renormalizability ensures that the mass function does not depend on the renormalization point. For the simplicity of notation, we will omit displaying the $\mu$ dependence altogether.
The general form of the gluon propagator is:
\begin{equation}
D_{\mu\nu}(q) = \frac{D(q^{2})}{q^{2}}\left[\delta_{\mu\nu}-\frac{q_{\mu} \; q_{\nu}}{q^{2}}\right]+\xi \frac{q_{\mu} \;q_{\nu}}{q^{4}},
\end{equation}
where $D(q^2)$ is the gluon dressing function and $\xi$ is the covariant gauge parameter. Due to the corresponding Ward identity, the longitudinal term proportional to $\xi$ does not get corrections at any order of perturbation theory. Hence $\xi=0$ is an obvious first option to work with. It corresponds to the Landau gauge, which is a convenient and natural choice for several reasons. Among others, model dependent differences between various \emph{ans\"atze} for the QGV are least noticeable in this gauge~\cite{Binosi:2016wcx}. Moreover, it is a covariant gauge which is readily implemented in lattice QCD simulations~\cite{Cucchieri:2009kk,Boucaud:2018xup}.

The first model of the gluon propagator employed herein is the MT interaction~\cite{Maris:1999nt}, where the effective coupling, $\alpha_{s}(q^2)\equiv g^2 D(q^2)/4\pi$, has the following form:
\begin{eqnarray}\label{eq:MTMODEL}
\hspace{-4mm} \frac{\alpha_s(q^2)}{q^2} = \frac{\pi D}{\omega^{6}}  q^{2}  e^{-q^{2}/\omega^{2}}
+\frac{ \gamma_{m} \; \pi \textit{F}(q^{2})}{\frac{1}{2}\ln[\tau+(1+q^{2}/\Lambda_{QCD}^{2})^{2}]},
\end{eqnarray}
with $\textit{F}(q^{2}) = \{1-e^{-q^{2}/[4 m_{t}^{2}]}\}/q^{2}$, $\tau = e^{2}-1$, $\gamma_{m} = 12/(33-2 N_{f})$, $N_{f} = 4$,
$m_t = 0.5$ GeV and $\Lambda_{QCD} = 0.234$ GeV.
 The first term provides an infrared enhancement, controlled by the parameters $\omega$ and $D$, while the second term reproduces the one-loop renormalization group equation of QCD. 
 
 The other model choice is the QC interaction~\cite{Qin:2011dd}:
 \begin{eqnarray}\label{eq:QCMODEL}
\hspace{-4mm} \frac{\alpha_s(q^2)}{q^2} = \frac{2\pi D}{\omega^{4}}  e^{-q^{2}/\omega^{2}}
+\frac{ \gamma_{m} \; \pi \textit{F}(q^{2})}{\frac{1}{2}\ln[\tau+(1+q^{2}/\Lambda_{QCD}^{2})^{2}]},
\end{eqnarray}
 which differs from the MT model in the infrared enhancement term. It ensures the behavior of the effective gluon is in agreement with our modern understanding of QCD's gauge sector; 
 in the minimal Landau gauge in 3+1-dimensions, the gluon propagator is a bounded, regular function
of spacelike momenta and is infrared enhanced~\cite{Kern:2019nzx}. 
 It has been confirmed that the hadron properties are insensitive to small variations of $\omega \in [0.4,0.6]$, so long as the product $m_G^3 \equiv (\omega D)$ remains constant (typical values of $m_G \sim 0.4 - 0.8$ GeV)~\cite{Chang:2013pq,Gao:2017uox}. These models have been extensively employed to study hadron physics through the SDEs of QCD, obtaining a wide range of predictions: meson and baryon spectrum and properties~\cite{Eichmann:2016yit,Qin:2018dqp,Qin:2019hgk}, parton distribution amplitudes~\cite{Chang:2013pq}, parton distribution functions~\cite{Nguyen:2011jy,Ding:2019lwe}, electromagnetic elastic~\cite{Chang:2013nia} and transition form factors~\cite{Raya:2015gva,Eichmann:2017wil,Raya:2016yuj,Ding:2018xwy}, hadronic contribution to the anomalous electromagnetic moment of the muon\cite{Raya:2019dnh,Eichmann:2019bqf,Eichmann:2019tjk,Aoyama:2020ynm}, etc.

\section{Quark-gluon vertex}
\label{sec:vertices}
 For the 1PI QGV, the simplest choice is to replace
 the fully dressed fermion-boson vertex by its tree level counterpart. Along with the ladder approximation of the meson Bethe-Salpeter equation, it corresponds to the rainbow-ladder truncation. Even in the abelian case of QED, this (bare) vertex manages to satisfy the corresponding WGFTI~\cite{Ward:1950xp,Green:1953te,Fradkin:1955jr,Takahashi:1957xn} 
 only in the chirally symmetric phase in the Landau gauge and in the leading log approximation for the wave-function renormalization~\cite{Bashir:1994az}. For these particular choices and limits, it ensures $Z(p^2;\mu^2)=1$ and $M(p^2)=0$. The simplicity of this choice brings even further undesirable features. It obviously lacks all those six basis structures which are dynamically generated through
 DCSB. Moreover, the associated dressed quark anomalous chromomagnetic moment and electromagnetic {\em distribution} in the infrared, associated with DCSB, is much less than what is required from observed hadron phenomenology~\cite{Chang:2010hb}. Some of these drawbacks can be compensated by a proper choice of parameters in the effective gluon propagator to render good description of pseudoscalar and light vector meson spectra~\cite{Chang:2011ei}. However, this is not the case with axial vector mesons, since the bare vertex lacks a proper enhancement of the spin-orbit splitting in this channel.

 In constructing a fully consistent fermion-boson vertex {\em Ansatz}, many efforts have been made over the last few decades. We choose to explore~\cite{Ball:1980ay,Curtis:1990zs,Kizilersu:2009kg,Bashir:2011dp} for our numerical investigation. The general form of the vertex consists of 12 linearly independent structures, which can be obtained from three vectors $k_{\mu}$, $p_{\mu}$, $\gamma_{\mu}$ and four spin scalars 1, $\gamma \cdot k$, $\gamma \cdot p$, $\gamma \cdot k \gamma \cdot p$. A first step towards constructing a QGV is employing the STI~\cite{Slavnov:1972fg,Taylor:1971ff}. In addition to the quark propagator, it also involves the ghost propagator and the quark-quark-ghost-ghost scattering kernel.  
 As we work with infrared enhanced effective one gluon exchange models, we can adopt the Abelian approximation of the STI with impunity, namely, the WGFTI which entails:
\begin{eqnarray*}
\label{eq:WGTI}
iq_{\mu} \Gamma_{\mu} = S^{-1}(k)-S^{-1}(p) \;.
\end{eqnarray*}
Following the usual arguments of Ball and Chiu, WGFTI allows this vertex to be decomposed into a longitudinal and a transverse part
\begin{eqnarray*}
\Gamma_{\mu}(k,p) = \Gamma^{L}_{\mu}(k,p)+\Gamma^{T}_{\mu}(k,p)\;,
\end{eqnarray*}
such that  $q_{\mu}\Gamma^{T}_{\mu} = 0$ and the longitudinal term ($\Gamma_\mu^L$) is fixed by the above WGFTI. In the so-called Ball-Chiu basis, $\Gamma_\mu^L$ is written as:
\begin{eqnarray}
\nonumber
\Gamma_{\mu}^{L(BC)}(k,p) &=& \lambda_{1}(k^{2},p^{2}) \; \gamma_{\mu} - i\lambda_{2}(k^{2},p^{2}) \, t_{\mu}\\
\nonumber
&+& \lambda_{3}(k^{2},p^{2}) \, t_{\mu}  \gamma \cdot t /2 \\
\label{eq:BCV}
&+& \lambda_{4}(k^{2},p^{2}) \, t_{\nu} \sigma_{\mu\nu},
\end{eqnarray}
where the dressing functions are
\begin{eqnarray}
\nonumber
\lambda_{1}(k^{2},p^{2}) &=&  {\overline{\Delta}}_A(k^2,p^2)  \;, \\
\nonumber
\lambda_{2}(k^{2},p^{2}) &=& \Delta_B(k^2,p^2)  \;, \\
\nonumber
\lambda_{3}(k^{2},p^{2}) &=& \Delta_A(k^2,p^2) \;, \\
\lambda_{4}(k^{2},p^{2}) &=& 0\;,
\end{eqnarray}
 with $t=k+p$, $(k^2-p^2) \Delta_{\varphi}(k^2,p^2) \equiv \varphi(k^2) - \varphi(p^2)$ and $2 {\overline{\Delta}}_{\varphi}(k^2,p^2) \equiv \varphi(k^2) + \varphi(p^2)$. A generalization of the BC vertex to the non-Abelian case can be found in~\cite{Aguilar:2018epe}. 
Note that although $\lambda_{4}=0$ for QED, the contribution coming from the triple gluon vertex in QCD ensures that it is non-zero for the latter case~\cite{Davydychev:2000rt}. However, as we work with the abelianized version of QCD, we will stick to $\lambda_{4}=0$.
Notice also that $\lambda_2$ carries an explicit dependence on the mass function. It implies that its appearance in the chiral limit owes itself entirely to DCSB. In the following, we will take $\Gamma_\mu^L = \Gamma_\mu^{L(BC)}$ and discuss different choices of $\Gamma_\mu^T$.

The transverse part is decomposed as a linear combination of the 8 basis vectors $T_{i\mu}$, that is
\begin{eqnarray*}
\Gamma^{T}_{\mu} = \sum_{i=1}^{8} \tau_{i}(k^{2},p^{2},q^{2}) \hspace{0.5mm} T_{i\mu}(k,p)\;,
\end{eqnarray*}
where $\tau_{i}$ are unknown scalar functions. Rather generally, the basis vectors can be written as:
\begin{eqnarray}
\nonumber
T_{1\mu}(k,p) &=& i\left[ p_{\mu}(k\cdot q)-k_{\mu}(p\cdot q) \right] \;, \\
\nonumber
T_{2\mu}(k,p) &=& [p_{\mu}(k\cdot q)-k_{\mu}(p\cdot q)] \,  \gamma \cdot t \;, \\
\nonumber
T_{3\mu}(k,p) &=& q^{2}\gamma_{\mu}-q_{\mu} \, \gamma \cdot q  \;, \\
\nonumber
T_{4\mu}(k,p) &=&iq^{2}[\gamma_{\mu}  \gamma \cdot t - t_{\mu}]
+ 2  q_{\mu}  p_{\nu} k_\lambda \sigma_{\nu\lambda}\;, \\
\nonumber
T_{5\mu}(k,p) &=&   \sigma_{\mu \nu} q_{\nu}\;, \\
\nonumber
T_{6\mu}(k,p) &=& \gamma_{\mu}  (p^{2}-k^{2}) +  t_{\mu} \, \gamma \cdot q \;, \\
\nonumber
T_{7\mu}(k,p) &=& \frac{i}{2}  (k^{2}-p^{2}) [\gamma_{\mu}  \gamma \cdot t - t_{\mu}] 
+  t_{\mu}  p_{\nu} k_{\lambda}    \sigma_{\nu\lambda}\;, \\
T_{8\mu}(k,p) &=& - i \gamma_{\mu}  p_{\nu} k_{\lambda}  \sigma_{\nu\lambda}+  k_{\mu}  \gamma \cdot p-p_{\mu} \gamma \cdot k \;.
\end{eqnarray}
Note that 
\begin{eqnarray}
q_{\mu} T_{i\mu}(k,p)=0  \qquad i=1,\cdots,8.  
\end{eqnarray}
This basis is not the one employed in~\cite{Ball:1980ay}. We 
choose to work with a modification of this initial basis which was
put forward in~\cite{Kizilersu:1995iz} and later employed in~\cite{Davydychev:2000rt} 
as well. This latter, which we have adopted here, choice ensures all transverse form factors of
the vertex are independent of any kinematic singularities
in one-loop perturbation theory in an arbitrary covariant
gauge.

The determination of the coefficients $\tau_i$ is not arbitrary. To a reasonable extent, they are constrained by the TTI, LKFT, MR, freedom of kinematic singularities and the adequate perturbation theory limit in the weak coupling regime~\cite{Ball:1980ay,Qin:2013mta,Albino:2018ncl}. 

Curtis and Pennington~\cite{Curtis:1990zs} adopted a simple choice of the transverse coefficients, which ensures MR of the  massless electron propagator in the quenched approximation of quantum electrodynamics (QED). This transverse part of the vertex,
referred to as the CP vertex, is merely:
\begin{eqnarray}
\Gamma^{T(CP)}_{\mu} = \frac{\gamma_{\mu}  (k^{2}-p^{2})-t_{\mu}  \gamma \cdot t}{2 \; d(k,p)} [A(k^{2})-A(p^{2})],
\end{eqnarray}
where $t=k+p$ and 
\begin{eqnarray*}
 d(k,p) &=& \frac{(k^{2}-p^{2})^{2}+\left[M^2(k^2)+ M^2(p^2)\right]^{2}}{k^{2}+p^{2}}\,.
\end{eqnarray*}
There is a peculiar $\left[M^2(k^2)+ M^2(p^2)\right]^{2}$ factor in this {\em Ansatz}.
Notice that its absence introduces an unwanted kinematic singularity. Moreover, 
it does not jeopardize the MR of the massless electron propagator by construction.

In a subsequent work, Kizilersu and Pennington~\cite{Kizilersu:2009kg} proposed two vertex constructions for the unquenched case, in the chiral limit with $n_f=1$. On using any of these two {\em Ans$\ddot{a}$tze} in the SDEs for the perturbative photon and massless fermion propagators simultaneously, they get the correct power law behavior for the photon dressing function and the fermion wave-function renormalization.
Both proposals satisfy the same constraints and differ only beyond the leading logarithmic order, while also giving similar results in the Landau gauge~\cite{Kizilersu:2009kg,Kizilersu:2013hea}. Thus, one could use either. We choose to work with the the following KP construction~:
\begin{equation}
\Gamma^{T(KP)}_{\mu} = \tau_{2}T_{2\mu}+\tau_{3}T_{3\mu}+\tau_{6}T_{6\mu}+\tau_{8}T_{8\mu},
\end{equation}
where the dressing functions are:
\begin{eqnarray}
\nonumber
\tau_{2}(k^{2},p^{2},q^{2}) &=& -\frac{4}{3}  \frac{1}{k^{4}-p^{4}}  (A(k^{2})-A(p^{2}))
\\
\nonumber
&-&\frac{1}{3}  \frac{A(k^{2})+A(p^{2})}{(k^{2}+p^{2})^{2}} \;   \ln\left[\frac{A(k^{2}) \;  A(p^{2})}{A(q^{2})^{2}}\right]\;, \\
\nonumber
\tau_{3}(k^{2},p^{2},q^{2}) &=& \frac{5}{12}\frac{1}{k^{2}-p^{2}}  (A(k^{2})-A(p^{2}))\\
\nonumber
&+&\frac{1}{6}\frac{A(k^{2})+A(p^{2})}{(k^{2}+p^{2})^{2}}  \;  \ln\left[\frac{A(k^{2}) \;  A(p^{2})}{A(q^{2})^{2}}\right]\;, \\ \nonumber
\tau_{6}(k^{2},p^{2},q^{2}) &=& -\frac{1}{4}\frac{1}{k^{2}+p^{2}}  (A(k^{2})-A(p^{2}))\;, \\
\tau_{8}(k^{2},p^{2},q^{2}) &=& 0\;.
\end{eqnarray}
In 2012, Bashir {\em et al.}~\cite{Bashir:2011dp} put forward a family of fermion-boson vertices expressed solely in terms of the vector and scalar functions appearing in the fermion propagator. Among other requirements, constraints on $a_i$ ensure the {\em Ansatz} is consistent with one-loop perturbation theory. For the sake of computational simplicity, the coefficients of the transverse basis are chosen to be independent of the angle between the relative momenta. Strikingly, it also has no explicit dependence on the covariant-gauge parameter.  Residual freedom of choice for $a_i$ allows us to achieve the gauge-independence of the critical coupling in QED, above which chiral symmetry is dynamically broken. The set of scalar functions $\tau_i$ for this proposal are written as:
\begin{align}\label{eq:2.69}
\tau_{1}(k,p) &= \frac{a_{1} \Delta_{B}(k^{2},p^{2})}{(k^{2} + p^{2})}\;,    \nonumber \\
\tau_{2}(k,p) &= \frac{a_{2} \Delta_{A}(k^{2},p^{2})}{(k^{2} + p^{2})}\;,    \nonumber \\
\tau_{3}(k,p) &= a_{3} \Delta_{A}(k^{2},p^{2})\;,    \nonumber \\
\tau_{4}(k,p) &= \frac{a_{4} \Delta_{B}(k^{2},p^{2}) }{[k^2 + M^2(k^2)][p^2 + M^2(p^2)]} \frac{k^2-p^2}{4}\;,    \nonumber \\
\tau_{5}(k,p) &= +a_{5} \Delta_{B}(k^{2},p^{2})\;,    \nonumber \\
\tau_{6}(k,p) &= -\frac{a_{6} (k^4 - p^4) \Delta_{A}(k^{2},p^{2}) }{[(k^2 - p^2)^{2} + (M^2(k^2) + M^2(p^2))^{2}]}\;,    \nonumber \\
\tau_{7}(k,p) &= +\left[\frac{a_7}{(k^{2} + p^{2})} +\frac{2(k-p)^2}{k^2-p^2}\tau_4\right]\Delta_{B}(k^2,p^2)\;,    \nonumber \\
\tau_{8}(k,p) &= a_{8} \Delta_{A}(k^{2},p^{2})\;,
\end{align}
where $a_i$ are momentum-independent constants whose values are listed in Table~\ref{table:tab21}. Such constants are interconnected by numerous constraints from perturbation theory and gauge covariance. The fixing procedure can be found in Ref.~\cite{Bashir:2011dp,Albino:2018ncl}. We call this proposal the BB vertex. In the next section we shall discuss the gap equation with all these vertices.
\begin{table}
\begin{ruledtabular}
\begin{tabular}{lcccccccc}
Constant & $a_{1}$ & $a_{2}$ & $a_{3}$  & $a_{4}$ & $a_{5}$ & $a_{6}$ & $a_{7}$ & $a_{8}$ \\
Value & 0 & 3.4 & 1  &  1 & -4/3 & -1/2 & 2.167  & -3.7 \\
\end{tabular}
\end{ruledtabular}
\caption{\label{table:tab21} A choice of momentum and gauge-independent coefficients of the transverse basis in the BB fermion-boson vertex~\cite{Bashir:2011dp,Albino:2018ncl}.}
\end{table}

\section{Gap equation}
\label{sec:gapequation}
Dressing functions $B(k^{2})$ and $A(k^{2})$ can be decoupled through proper projections of Eq. \eqref{eq:1}, \emph{viz.}, multiplying  Eq. \eqref{eq:1}  by $\mathbb{1}$ and $\slashed{p},$ respectively, and then taking traces. \\ \\
\noindent
{\bf{The bare vertex:}} For this approximation $\Gamma_{\mu}  = \gamma_{\mu}$, quark self-energy of Eq. \eqref{eq:2} acquires the following simple form
\begin{equation}\label{eq:bare}
\Sigma(p) = Z_{1}C_{F}\int_k  \;  g^2  D_{\mu\nu}(q) \gamma_{\mu} \;  S(k) (Z_2 \gamma_{\nu})\;.
\end{equation}
Using the steps suggested above, one arrives at the expressions:
\begin{eqnarray}
\label{intbare1}
B(p^{2}) &=& m_{0}  Z_2+16 \pi Z_2^2\int_k \frac{\alpha_{s}(q^{2})}{q^{2}} \sigma_s(k^2)
\;,\\
\nonumber
A(p^{2}) &=& Z_2+\frac{16 \pi}{3 \;  p^{2}}Z_2^2\int_k \frac{\alpha_{s}(q^{2})}{q^{2}}
\\
\label{intbare2}
&\times&  \sigma_v(k^2) \left[k\cdot p+\frac{2 \;  k\cdot q \; p\cdot q}{q^{2}}\right]\;.
\end{eqnarray}
This minimal {\em Ansatz} neglects any non-Abelian contribution to the QGV Therefore, for the sake of consistency, we equate $Z_1 = Z_2$. In fact, we can continue to use it with impunity for the BC, CP, KP and BB vertices as they were all proposed in an Abelian set up. Note that this reasoning is no longer valid if the QGV employed is constructed form the corresponding STI because this extended identity incorporates the effects coming from the non-Abelian ghost-gluon sector. However, note that the renormalization boundary condition, independently of the truncation, entails
\begin{eqnarray*}
S^{-1}(p)|_{p^2=\mu^2}=i \gamma \cdot p+m(\mu)\;,
\end{eqnarray*}
 where $m(\mu)=M(\mu^2)$ is the scale dependent running quark mass. The above condition implies $A(p^2=\mu^2)=1$ and 
 $B(p^2=\mu^2)= m(\mu)$. One can thus define a convenient renormalization point invariant mass as follows:
 \begin{equation}
    \label{eq:RGImass}
     \hat{m} = m(\mu) \left[ \frac{1}{2} \ln{\left(\frac{\mu^2}{\Lambda_{QCD}^2}\right)} \right]^{\gamma_m}\;.
 \end{equation}
{\bf{The BC vertex:}} Returning to the gap equation and the QGV, if one employs the BC vertex, Eq. \eqref{intbare1} modifies as:
\begin{eqnarray}
\nonumber
B(p^{2}) &=& m_{0} Z_2+\frac{16\pi}{3}Z_2\int_k\;\frac{\alpha_{s}(q^{2})}{q^{2}} \\
\label{eq:23}
&\times& \left\{ \sigma_s(k^2)[I_{B1}^{BC}+I_{B2}^{BC}] + \sigma_v(k^2)[I_{B3}^{BC}] \right\},
\end{eqnarray}
where $I_{B1}^{BC}$, $I_{B2}^{BC}$ and $I_{B3}^{BC}$ are the integrands related to the BC vertex, such that
\begin{eqnarray*}
 I_{B1}^{BC} &=& 3 \frac{A(k^{2})+A(p^{2})}{2}, \\
 I_{B2}^{BC} &=&    \Delta_A(k^2,p^2)
  \left\{ \frac{t^{2} \; q^{2}- ( t\cdot q )^{2}}{ 2q^{2}}\right\}, \\
 I_{B3}^{BC} &=&  \Delta_B(k^2,p^2)  
 \left\{ \frac{q^{2}t \cdot k-t \cdot q \; q\cdot k}{q^{2}}\right\}.
\end{eqnarray*}
Analogously, the corresponding equation for $A(p^2)$ is:
\begin{eqnarray}
\nonumber
A(p^{2}) &=& Z_2+\frac{16  \pi}{3}Z_2\int_k\; \frac{\alpha_{s}(q^{2})}{q^{2}}
\\
\label{eq:24}
&\times& \left\{\sigma_v(k^2)[I_{A1}^{BC}-I_{A2}^{BC}]+\sigma_s(k^2)[I_{A3}^{BC}] \right\},
\end{eqnarray}
where the integrands are written as
\begin{eqnarray*}
I_{A1}^{BC} &=& \frac{A(k^{2})+A(p^{2})}{2}\frac{1}{p^{2}}
\\
&\times& \left\{\frac{k\cdot p\; q^{2}+2 \; [(k^{2}+p^{2}) \; k\cdot p-k^{2} \; p^{2}-k\cdot p^{2}]}{q^{2}} \right\}\;, \\
\nonumber
I_{A2}^{BC} &=& \frac{1}{2p^{2}} \Delta_A(k^2,p^2) 
\\
&\times& 
\bigg\{[p^2 k + k^2 p] \cdot t
- \frac{p^{2} t\cdot q \; k\cdot q - k^{2} t \cdot q \; p\cdot q}{q^{2}} \bigg\}\;,\\
\nonumber
I_{A3}^{BC} &=& \Delta_B(k^2,p^2)  \frac{1}{p^{2}}  \left\{\frac{t\cdot q \; p\cdot q-t\cdot p \; q^{2}}{q^{2}} \right\}\;.
\end{eqnarray*}
{\bf{The CP-vertex:}} By taking into account the transverse CP vertex, one arrives at:
\begin{eqnarray}
\nonumber
B(p^{2}) &=& m_{0}Z_2+\frac{16\pi}{3}Z_2\int_k\;\frac{\alpha_{s}(q^{2})}{q^{2}} \\
\nonumber
&\times&
\Big\{ \sigma_s(k^2)[I_{B1}^{BC}+I_{B2}^{BC}] + \sigma_v(k^2)[I_{B3}^{BC}] \\
&+& \frac{3}{2}\sigma_s(k^2)(k^2+p^2) L(k^2+p^2) \Big\}\;,
\end{eqnarray}
where $L\equiv L(k^{2},p^{2})$ is defined as
\begin{eqnarray}
\nonumber
L =
\frac{[A^{2}(k^{2})A^{2}(p^{2})]^{2} 
\Delta_A(k^2,p^2) }{[A^{2}(k^{2})A^{2}(p^{2})]^{2}+[A^{2}(p^{2})B^{2}(k^{2})+A^{2}(k^{2})B^{2}(p^{2})]^{2}}. \quad
\nonumber
\end{eqnarray}
On the other hand, the analogous expression for $A(p^2)$ reads as:
\begin{eqnarray}
\nonumber
A(p^{2}) &=& Z_2+\frac{16 \pi}{3}Z_2 \int_k\;\frac{\alpha_{s}(q^{2})}{q^{2}}
\\
\nonumber
&\times& \Big\{ \sigma_v(k^2)[I_{A1}^{BC}-I_{A2}^{BC}]+\sigma_s(k^2)[I_{A3}^{BC}] \\
&+& 2 \sigma_v(k^2) \frac{k^2+p^2}{k^2-p^2}[I_{A1}^{CP}+I_{A2}^{CP}] L(k^2,p^2) \Big\}\;.
\end{eqnarray}
The integrands $I_{A1}^{CP}$ and $I_{A2}^{CP}$, related to the CP term, are
\begin{eqnarray*}
I_{A1}^{CP} &=& (k^{2}-p^{2})\left\{\frac{3(k^{2}+p^{2})k\cdot p-2 k^{2}  p^{2}-4k\cdot p^{2}}{q^{2}}\right\}, \\
I_{A2}^{CP} &=& k^{2} t \cdot p -p^{2} t\cdot k  + \frac{p^{2} t \cdot q k\cdot q - k^{2} t \cdot q  p\cdot q}{q^{2}}. \nonumber
\end{eqnarray*}
{\bf{The KP vertex:}} The KP vertex {\em Ansatz} yields the following equation for $B(p^{2})$:
\begin{eqnarray}
\nonumber
B(p^{2}) &=& m_{0}Z_2+\frac{16  \pi}{3}Z_2\int_k\;\frac{\alpha_{s}(q^{2})}{q^{2}}
\\
\nonumber
&\times& \Big\{ \sigma_s(k^2)[I_{B1}^{BC}+I_{B2}^{BC}] + \sigma_v(k^2)[I_{B3}^{BC}] 
\\
&+& \sigma_s(k^2)[I_{B1}^{KP}-I_{B2}^{KP}-I_{B3}^{KP}] \Big\}\;.
\end{eqnarray}
The integrands related specifically to the KP vertex, $I_{B1}^{KP}$, $I_{B2}^{KP}$ and $I_{B3}^{KP}$, can be expressed as
\begin{eqnarray*}
I_{B1}^{KP} &=& 2 \;  (k\cdot p^{2}-k^{2}p^{2})\bigg\{\frac{4}{3}\frac{A(k^{2})-A(p^{2})}{k^{4}-p^{4}}
\\
&+&\frac{1}{3}\frac{A(k^{2})+A(p^{2})}{(k^{2}+p^{2})^{2}} \;  \ln\left[\frac{A(k^{2}) \; A(p^{2})}{A^{2}(q^{2})}\right]\bigg\}\;, \\
I_{B2}^{KP} &=& q^{2}\bigg\{\frac{5}{4} \Delta_A(k^2,p^2) \;,
\\
&+& \frac{1}{2}\frac{A(k^{2})+A(p^{2})}{k^{2}+p^{2}}\ln\left[\frac{A(k^{2}) \; A(p^{2})}{A^{2}(q^{2})}\right]\bigg\}\;, \\
I_{B3}^{KP} &=& \frac{3}{4}(k^{2}-p^{2})\frac{A(k^{2})-A(p^{2})}{k^{2}+p^{2}}\;.
\end{eqnarray*}
The corresponding equation for $A(p^2)$, for KP vertex, is:
\begin{eqnarray}
\nonumber
A(p^{2}) &=& Z_2+Z_2 \frac{16 \pi}{3}\int_k\;\frac{\alpha_{s}(q^{2})}{q^{2}}
\\
\nonumber
&\times& \Big\{ \sigma_v(k^2)[I_{A1}^{BC}-I_{A2}^{BC}]+\sigma_s(k^2)[I_{A3}^{BC}]
\\
&+&\sigma_v(k^{2})(I_{A1}^{KP}+I_{A2}^{KP}-I_{A3}^{KP}) \Big\}\;,
\end{eqnarray}
where the related integrands are
\begin{eqnarray*}
I_{A1}^{KP} &=& \frac{(k^{2}+p^{2})(k^{2} \;  p^{2}-k.p^{2})}{p^{2}}\bigg\{\frac{4}{3}\frac{A(k^{2})-A(p^{2})}{k^{4}-p^{4}}
\\
\nonumber
&+&\frac{1}{3}\frac{A(k^{2})+A(p^{2})}{(k^{2}+p^{2})^{2}} \;  \ln\left[\frac{A(k^{2}) \;  A(p^{2})}{A^{2}(q^{2})}\right]\bigg\}, \\
\nonumber
I_{A2}^{KP} &=& \frac{k\cdot p \; (4 \; k\cdot p-3 \; p^{2})+k^{2} \; (2 \; p^{2}-3 \; k\cdot p)}{p^{2}} \\
\nonumber
&\times& \bigg\{\frac{5}{12} \Delta_A(k^2,p^2) \\
&+&\frac{1}{6}\frac{A(k^{2})+A(p^{2})}{k^{2}+p^{2}} \; \ln\left[\frac{A(k^{2}) \; A(p^{2})}{A^{2}(q^{2})}\right]\bigg\},
\\
I_{A3}^{KP} &=& \frac{3}{4} \;  A(k^{2}) \;  [A(k^{2})-A(p^{2})]\frac{(k^{2}-p^{2}) \;  }{(k^{2}+p^{2})}\frac{k\cdot p}{p^{2} }\;.
\end{eqnarray*}
Unlike the other vertex \emph{Ans\"atze}, the transverse part of the KP vertex introduces a non-trivial angular dependence, in connection to the logarithmic terms which contain $A(q^2)$. Thus, the numerical evaluation of such integrals is considerably more complicated. 

\noindent
{\bf{The BB vertex:}} Finally, the integral equations for the scalar functions $B(p^2)$ and $A(p^2)$, using the
 BB vertex,~\cite{Bashir:2011dp} are written as:
\begin{align}
B(p^2) &= \mbox{r.h.s. of Eq. } \eqref{eq:23} \nonumber \\
       &- \frac{16\pi}{3} Z_2 \int_k\; \frac{\alpha_{s}(q^2)}{q^2} \frac{1}{k^2 A^2(k^2)+ B^2(k^2)}      \nonumber \\
       & \times \bigg\{ A(k^2) ( (k \cdot p)^2-k^2p^2 ) \tau_{1}  \nonumber \\
       & + 2 B(k^2) ( (k \cdot p)^2-k^2p^2 ) \tau_{2}       \nonumber \\
       & -3 B(k^2) (k^2 + p^2 - 2 k\cdot p)  \tau_{3}  \nonumber \\
       & + A(k^2) [k^2(p^2-3k \cdot p)\nonumber \\
       & +k \cdot p(3p^2-4 k \cdot p)+3k^4 ] \tau_{4} \nonumber\\
       & +3 A(k^2) (k^2-k\cdot p) \tau_{5}  \nonumber \\
       & + 3 B(k^2) (k^2-p^2) \tau_{6}  \nonumber \\
       & + \frac{A(k^2)}{2}  [k^2(3k\cdot p-p^2) \nonumber \\
       & -k\cdot p(2k\cdot p+3p^2) + 3k^4)] \tau_{7} \nonumber\\
       & + 3 a_{8} B(k^2) k\cdot p \Delta_A(k^2,p^2)     \bigg\} ,
\end{align}

\begin{align}
A(p^2) &= \mbox{r.h.s. of Eq. }   \eqref{eq:24}  \nonumber \\
       &- \frac{16\pi}{3}Z_2 \int_k \frac{\alpha_{s}(q^2)}{q^2} \frac{1}{k^2 A^2(k^2)+ B^2(k^2)}     \nonumber \\
       &\times \frac{1}{p^2} \bigg\{  B(k^2) ((k\cdot p)^2 - k^2 p^2) \tau_{1} \nonumber \\
       & +  A(k^2) (k^2 (p^4-(k\cdot p)^2)-p^2(k\cdot p)^2+k^4p^2) \tau_{2}       \nonumber \\
       & -  A(k^2) (k^2(3 k\cdot p-2 p^2)+k\cdot p(3p^2-4k\cdot p)) \tau_{3}  \nonumber \\
       & -  B(k^2) [-4(k\cdot p)^2-3p^2(k\cdot p)\nonumber \\
       & +k^2(3k\cdot p+p^2)+3p^4] \tau_{4} \nonumber \\
       & + 3  B(k^2) (p^2-k\cdot p) \tau_{5}            \nonumber \\
       & + 3  A(k^2) (k^2-p^2)(k\cdot p)  \tau_{6} \nonumber \\
       & - \frac{B(k^2)}{2}  [2(k\cdot p)^2-3p^2(k\cdot p) \nonumber \\
       & +k^2(3k \cdot p+p^2)-3p^4 ]   \tau_{7}      \nonumber \\
       & - 2  A(k^2) ((k\cdot p)^2-k^2 p^2) \tau_{8}      \bigg\} .
\end{align}
In the next section, we present and discuss the numerical results obtained from employing different vertex \emph{Ans\"atze} we chose to study the gap equation with.

\section{Results}
\label{sec:results}
We solve the gap equation for a few largely employed quark-gluon vertices suggested in literature over the last three decades. This exercise is carried out in conjunction with effective MT and QC models for the gluon propagator. The renormalization point is set to $\mu\equiv \mu_3=2.86$ GeV. The infrared strength of the gluon models is fixed from the chiral quark condensate~\cite{Williams:2006vva}, as we now explain. Note that the chiral limit is defined when $\hat{m}=0$ ($\mu \to \infty$) in Eq. \eqref{eq:RGImass} and we label it as $m_q$.  In this limit, we can express the chiral quark condensate as follows:
\begin{equation}
    \label{eq:condensate}
    -\textless\bar{q}q\textgreater_\mu^0 = Z_4 N_c \mbox{Tr} \int_k S_{\hat{m}}(k;\mu) \;.
\end{equation}
We can thus define a renormalization point invariant condensate $ \langle \bar{q}q \rangle $ as
\begin{equation}
\label{eq:condensate2}
    m(\mu) \textless\bar{q}q \textgreater_{\mu}^0 = \hat{m} \textless\bar{q}q \textgreater \;.
\end{equation}

\begin{table}
\begin{ruledtabular}
\begin{tabular}{lcccc}
 Vertex &  $\omega$ & $m_G$ &  $\omega$ & $m_G$ \\
\hline \\
& \multicolumn{2}{c}{MT Model~\cite{Maris:1999nt}}  & \multicolumn{2}{c}{QC Model~\cite{Qin:2011dd}} \\
\hline 
Bare & 0.4 & 0.728 & 0.4 & 0.744 \\
BC~\cite{Ball:1980ay} & 0.4 & 0.528 & 0.48 & 0.583\\
CP~\cite{Curtis:1990zs} & 0.4 & 0.516 & 0.4 & 0.529 \\
KP~\cite{Kizilersu:2009kg} & 0.4 & 0.528 & 0.46 & 0.551\\
BB~\cite{Bashir:2011dp} & 0.4 & 0.544 & 0.4 & 0.562 \\
\end{tabular}
\end{ruledtabular}
\caption{\label{tab:modelparams} Gluon model parameters $\omega$ and $m_G = (\omega D)^{1/3}$ for each vertex {\em Ansatz}. Dimensioned quantities are expressed in GeV.}
\end{table}

\begin{figure}[ht!]
\centering
\includegraphics[width=0.45\textwidth]{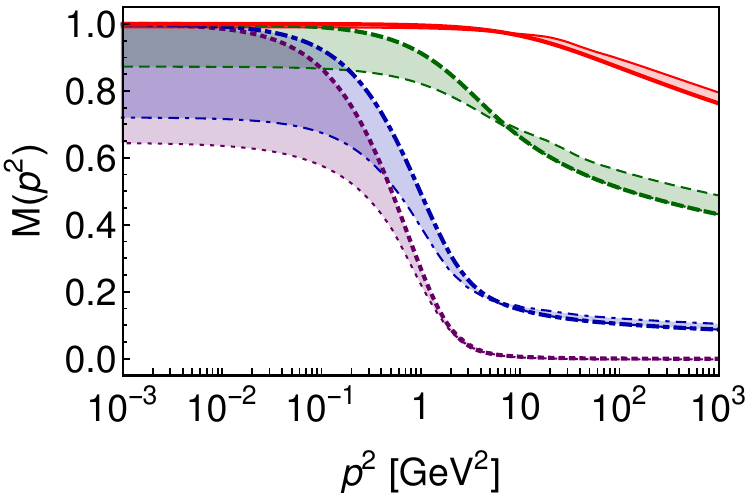}
        \caption{\textbf{[MT model]} Mass functions for different current quark masses and QGV \emph{Ans\"atze}. The boundaries of the bands are given by the lowest and highest produced values of $M(p^2)$, for the different vertices: BC, CP, KP, BB and the bare vertex. Purple band with dotted boundaries corresponds to the chiral limit. Blue (dot dashed boundary), green (dashed boundary) and red (solid boundary) bands correspond to $m_s = 0.1$ GeV, $m_c = 1$ GeV and $m_b = 4.1$ GeV, respectively. Mass functions or each quark flavor have been normalized such that $M_{\mbox{max}}(0) = 1$. Bare vertex results are highlighted with a thicker line which forms the top edge of each band.}\label{fig:MFMT1}
\end{figure}

\begin{figure}[ht!]
\centering
\includegraphics[width=0.45\textwidth]{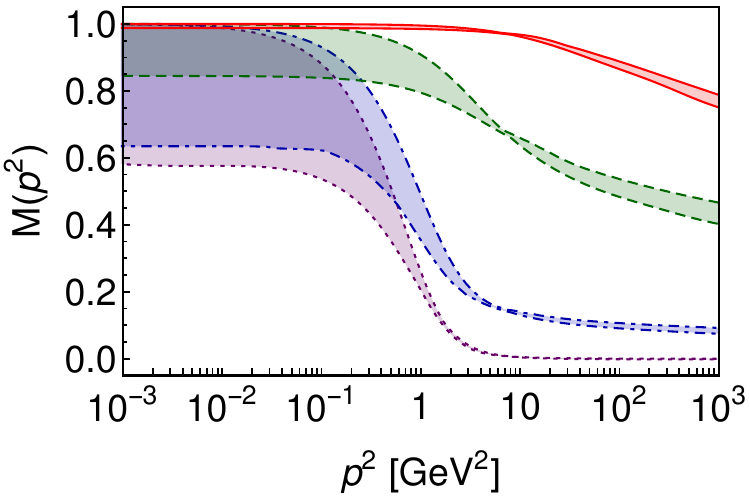}
        \caption{\textbf{[QC model]} Mass functions for different current quark masses and QGV \emph{Ans\"atze}. Bands and curves have the same meaning as in 
        Fig.~\ref{fig:MFMT1}.}\label{fig:MFQC1}
\end{figure}

\begin{figure}[ht!]
\centering
\includegraphics[width=0.45\textwidth]{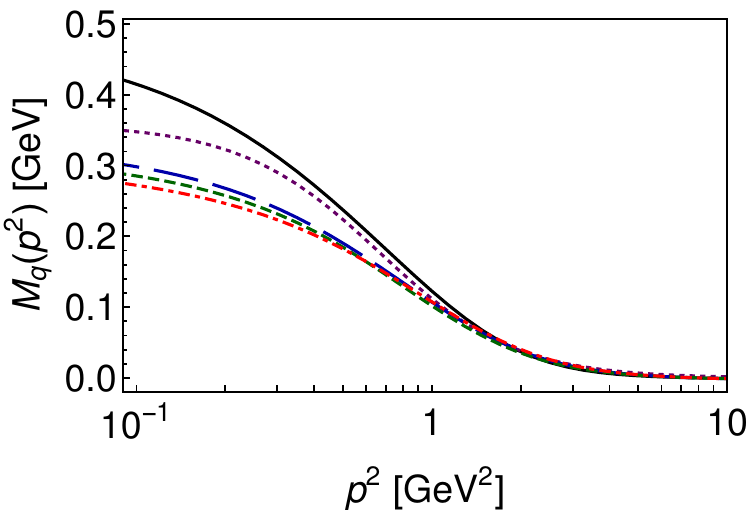}
        \caption{\textbf{[MT model]} Chiral limit mass functions from different QGV \emph{Ans$\ddot{a}$tze}. Bare (black, solid), Ball-Chiu (blue, long-dashed), Curtis-Pennington (green, dashed),  Kizilerzu-Pennington (Red, dot-dashed) and  Bashir-Bermudez (purple, dotted) vertex \emph{Ans\"atze}.}\label{fig:ChiralMT}
\end{figure}

\begin{figure}[ht!]
\centering
\includegraphics[width=0.45\textwidth]{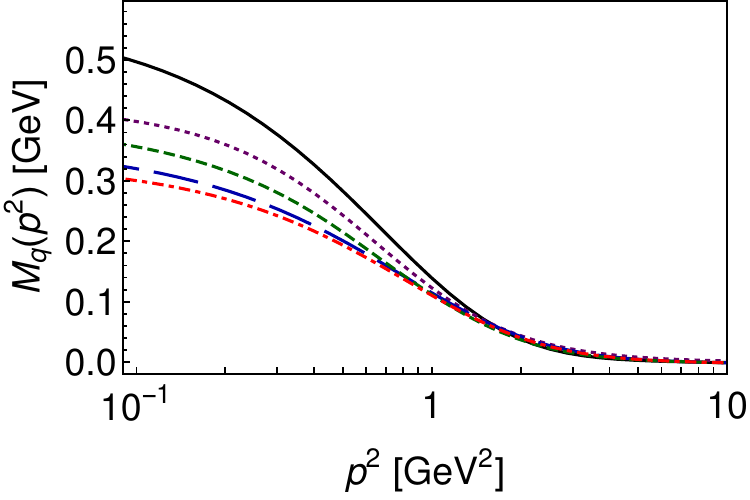}
        \caption{\textbf{[QC model]} Chiral limit mass functions from different QGV \emph{Ans\"atze}. Curves are labeled as in Fig.~\ref{fig:ChiralMT}.}\label{fig:ChiralQC}
\end{figure}

It is an order parameter of DCSB and, as explained elsewhere~\cite{Brodsky:2010xf,Chang:2011mu}, it is also the chiral limit value of the in-meson condensate. Therefore, it is natural to fix the effective gluon strength to produce a reasonable value of the chiral quark condensate and study its impact on other quantities. We fix MT and QC gluon model parameters, $\omega$ and the product $\omega D$, to obtain $-\textless\bar{q}q\textgreater_{\mu_3}^0 = (0.256$ GeV$)^3$. Along with Eq. \eqref{eq:condensate2}, this value yields: $-\textless \bar{q}q\textgreater_{\mu_2}^0 = (0.250$ GeV$)^3$ and  $- \textless \bar{q}q\textgreater_{\mu_{19}}^0 = (0.280$ GeV$)^3$ (where $\mu_{2,19} = 2,\;19$ GeV), in agreement with modern estimates~\cite{Ding:2018xwy}. The specific choice of parameters is displayed in 
Table~\ref{tab:modelparams}. The resulting chiral limit mass functions are shown in 
Figs.~\ref{fig:MFMT1}-\ref{fig:MFQC1}, along with those obtained for different non-zero current quark masses: $m(\mu) = 0.004,\;0.1,\;1.0,\;4.1$ GeV, labeled as $m_{u/d},\;m_s,\;m_c$ and $m_b$, respectively (the $m_{u/d}$ results are not displayed, in order to avoid overlap with the chiral limit results). In an intermediate range of momenta, Figs.~\ref{fig:ChiralMT}-\ref{fig:ChiralQC} show a more pronounced comparison of chiral limit results for different {\em Ans$\ddot{a}$tze} of the QGV.

The mass functions exhibit the expected features, namely: saturation at a finite value as $p^2\to 0$ and a monotonic decrease as $p^2$ increases. The saturation value, $M(0)$, in comparison with the current quark mass, is expectedly much larger in the case of the light quarks and it decreases sharply with increasing $p^2$, whereas it exhibits far less steep running for the heavy quarks. It is clear that dynamical mass generation via strong-interaction processes (DCSB) is the dominant mass generating mechanism in the light sector, while the heavy sector is largely overshadowed by its predominant coupling to the Higgs field.

\begin{figure}[ht!]
\centering
\includegraphics[width=0.45\textwidth]{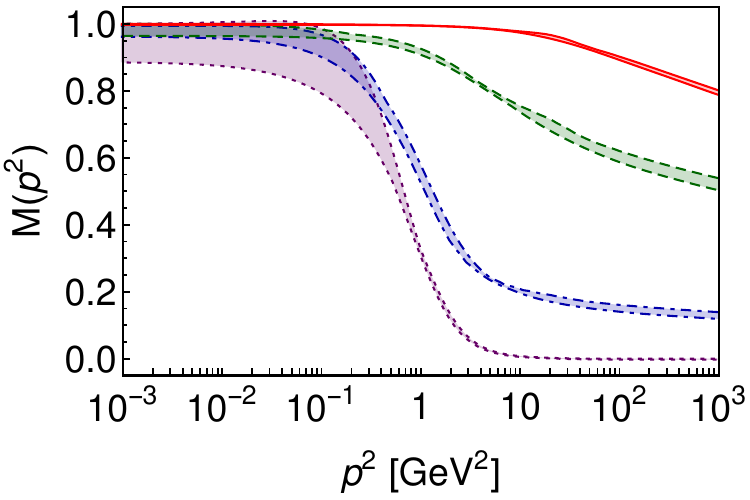}
        \caption{\textbf{[MT model]} Mass functions for different current quark masses and QGV \emph{Ans\"atze}. Bands and curves have the same meaning as in Fig.~\ref{fig:MFMT1}. However, the bare vertex results have been omitted in this figure.}\label{fig:MFMT2}
\end{figure}

\begin{figure}[ht!]
\centering
\includegraphics[width=0.45\textwidth]{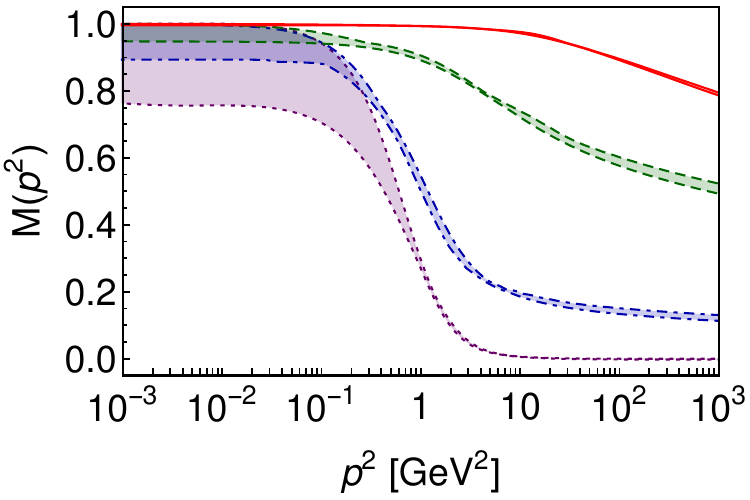}
        \caption{\textbf{[QC model]} Mass functions for different current quark masses and QGV \emph{Ans\"atze}. Bands and curves have the same meaning as in Fig.~\ref{fig:MFMT1}. The bare vertex results have again been excluded to bring out the robustness of the results for the dressed vertices.}\label{fig:MFQC2}
\end{figure}

Also readily observed is the fact that bare vertex calculations tend to produce larger values of $M(0)$ (although the asymptotic behavior of $M(p^2)$ for $p^2 {\rightarrow} \infty$ is reached faster). This is a consequence of the artificial enhancement of the effective coupling in order to produce sufficient amount of phenomenologically required DCSB\footnote{The strength needed to simultaneously produce reasonable values of vacuum quark condensate, mass spectrum and decay constants.}. In fact, if the infrared gluon model were enhanced just as much as with the other vertices, chiral condensate and $M(0)$ would decrease $40-60\%$. Moreover, had we omitted the bare vertex results, the bands in Figs.~\ref{fig:MFMT1}-\ref{fig:MFQC1} would have become much narrower. This is exactly what is displayed in Figs.~\ref{fig:MFMT2}-\ref{fig:MFQC2}. It clearly indicates that the results obtained from properly constructed quark-gluon vertices are more robust and less sensitive to the gluon models parameters. 

Another important feature of the mass function is its asymptotic behavior. In the chiral limit:~\cite{Roberts:1994dr,Williams:2006vva}:
\begin{equation}
\label{eq:MFA}
M(p^2\to \infty) \sim \frac{\ln{[p^2/\Lambda_{QCD}^2]}^{\gamma_m-1}}{p^2}\;.
\end{equation}
Naturally, since bare vertex is the leading order term in the perturbative expansion, mass function reaches this behavior faster in such case. It is followed (consistently with both MT and QC interactions) by CP, BC, KP and BB vertices, respectively. Beyond the chiral limit, Eq. \eqref{eq:MFA} is modified by including an extra term, which is proportional to the current quark mass~\cite{Roberts:1994dr,Williams:2006vva}, but the overall pattern persists.

\begin{figure}[ht!]
\centering
\includegraphics[width=0.45\textwidth]{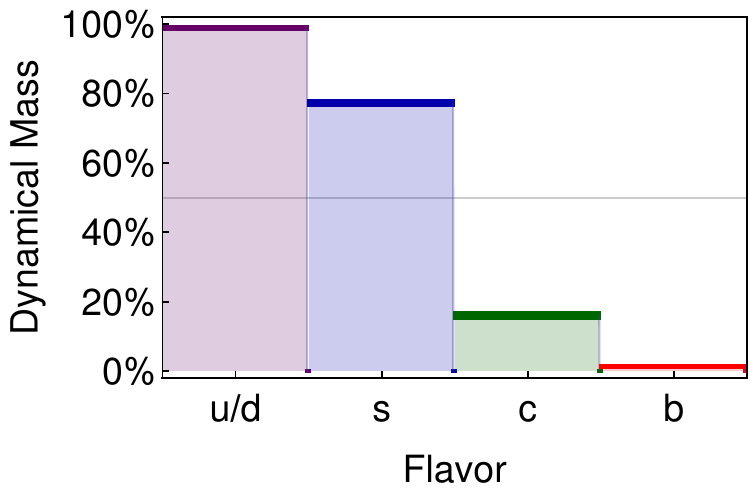}
\caption{Dynamical versus explicit mass generation as defined in Eq. \eqref{eq:DynVsExp}. The narrow darker regions correspond to the uncertainty coming from the choice of the QGV (BC, CP, KP, BB) and effective coupling (MT or QC). The horizontal line at the 50 \% mark corresponds to $m(\mu)=m_{\text{crit}} \approx 0.284$  GeV. It lies between the strange and charm quark masses. It corresponds to the ratio of Dynamical vs Explicit mass generation being $\bar{M}(m)=1/2$.}\label{fig:DynVsExp}
\end{figure}

To further understand the interplay between explicit and dynamical mass generation, let us define the following quantity:
\begin{equation}
\label{eq:DynVsExp}
    \bar{M}(m) = \left| \frac{M_E-m(\mu)}{M_E} \right|,
\end{equation}
where $M_E$, the constituent Euclidian mass, is defined through
\begin{eqnarray*}
M_E^2 \equiv \{ p^2\; | \; p^2 = M^2(p^2) \}\;.
\end{eqnarray*}
Naturally, $\bar{M}(m)\to1$ as $m(\mu)\to 0$, while  $\bar{M}(m)$ smoothly approaches zero with increasing current quark mass, as the explicit mass generation becomes dominant. Therefore, one can interpret the vicinity around $m_{crit}$, $m_{crit} \equiv \{ m(\mu) \; | \; \bar{M}(m) = 1/2 \}$, as the region in which the strengths of explicit and dynamical chiral symmetry breaking are comparable. We find that $m_{crit} \approx 0.284$ GeV, consistent with that obtained in~\cite{Serna:2018dwk}, from different criteria. Notice that $m_{crit}$ lies between charm and strange quark masses, but closer to the latter; thus, strange quark can be considered as the boundary between strong and weak mass generation mechanism being dominant~\cite{Ding:2018xwy}. Explicit values of $M(0)$ and $M_E$ for a set of current quark masses ($m_q,\;m_{u/d},\;m_s,\;m_c,\;m_b$) are listed in Tables~\ref{tab:m0}-\ref{tab:mE}. Figure~\ref{fig:DynVsExp} shows $\bar{M}(m)$ in a range of values of $m(\mu)$. Notably, $\bar{M}(m)$ is practically insensitive to the choice of the fully-dressed QGV (and gluon models), while those obtained from the bare vertex lie closely. Our observations are thus 
practically model independent statements.

\begin{table}

\begin{ruledtabular}
\begin{tabular}{lccccc}
Vertex & $ m_{q} $ & $ m_{u/d} $ & $ m_{s}$ & $ m_{c}$ & $ m_{b}$  \\
\hline \\
                      \multicolumn{6}{c}{MT Model~\cite{Maris:1999nt}}   \\
\hline
Bare                         & 0.484 & 0.492 & 0.649 & 1.464 & 4.228 \\
BC~\cite{Ball:1980ay}                 & 0.331 & 0.337 & 0.468 & 1.284 & 4.186  \\
CP~\cite{Curtis:1990zs}           & 0.315 & 0.323 & 0.468 & 1.279 & 4.184 \\
KP~\cite{Kizilersu:2009kg}        & 0.306 & 0.315 & 0.471 & 1.297 & 4.186 \\
BB~\cite{Bashir:2011dp}           & 0.353 & 0.356 & 0.483 & 1.186 & 4.072 \\
\hline 
&&& \\
                      \multicolumn{6}{c}{QC Model~\cite{Qin:2011dd}}  \\
\hline
Bare                        & 0.573 & 0.581 & 0.730 & 1.520 & 4.240 \\
BC~\cite{Ball:1980ay}                  & 0.360 & 0.366 & 0.485 & 1.295 & 4.188 \\
CP~\cite{Curtis:1990zs}          & 0.399 & 0.403 & 0.502 & 1.284 & 4.185 \\
KP~\cite{Kizilersu:2009kg}       & 0.330 & 0.339 & 0.479 & 1.296 & 4.187 \\
BB~\cite{Bashir:2011dp}            & 0.435 & 0.438 & 0.512 & 1.186 & 4.055 \\
\end{tabular}
\end{ruledtabular}
\caption{\label{tab:m0}Calculated constituent quark masses $M(0)$ for different current quark masses and QGV \emph{Ans\"atze}. Dimensioned quantities are expressed in GeV. Gluon model parameters are shown in Table~\ref{tab:modelparams}.}
\end{table}

\begin{table}

\begin{ruledtabular}
\begin{tabular}{lccccc}
Vertex & $ m_{q} $ & $ m_{u/d} $ & $ m_{s}$ & $ m_{c}$ & $ m_{b}$  \\
\hline \\
                    \multicolumn{6}{c}{MT Model~\cite{Maris:1999nt}} \\
\hline
Bare                         & 0.388 & 0.395 & 0.523 & 1.274 & 4.016 \\
BC~\cite{Ball:1980ay}                 & 0.301 & 0.307 & 0.417 & 1.184 & 4.037  \\
CP~\cite{Curtis:1990zs}           & 0.290 & 0.297 & 0.419 & 1.181 & 4.038 \\
KP~\cite{Kizilersu:2009kg}        & 0.280 & 0.287 & 0.430 & 1.198 & 4.036 \\
BB~\cite{Bashir:2011dp}           & 0.344 & 0.347 & 0.445 & 1.186 & 4.072 \\
\hline 
&&& \\
                    \multicolumn{6}{c}{QC Model~\cite{Maris:1999nt}} \\
\hline 
Bare                        & 0.442 & 0.449 & 0.574 & 1.303 & 4.009 \\
BC~\cite{Ball:1980ay}                  & 0.319 & 0.324 & 0.429 & 1.191 & 4.036 \\
CP~\cite{Curtis:1990zs}          & 0.347 & 0.352 & 0.439 & 1.183 & 4.038 \\
KP~\cite{Kizilersu:2009kg}       & 0.302 & 0.310 & 0.435 & 1.199 & 4.036 \\
BB~\cite{Bashir:2011dp}            & 0.390 & 0.391 & 0.449 & 1.173 & 4.055 \\
\end{tabular}
\end{ruledtabular}
\caption{\label{tab:mE}Calculated Euclidean constituent quark masses $M_E$ for different current quark masses and QGV \emph{Ans\"atze}. Dimensioned quantities are expressed in GeV. Gluon model parameters are shown in Table~\ref{tab:modelparams}.}
\end{table}

\begin{figure}[ht!]
\centering
\includegraphics[width=0.45\textwidth]{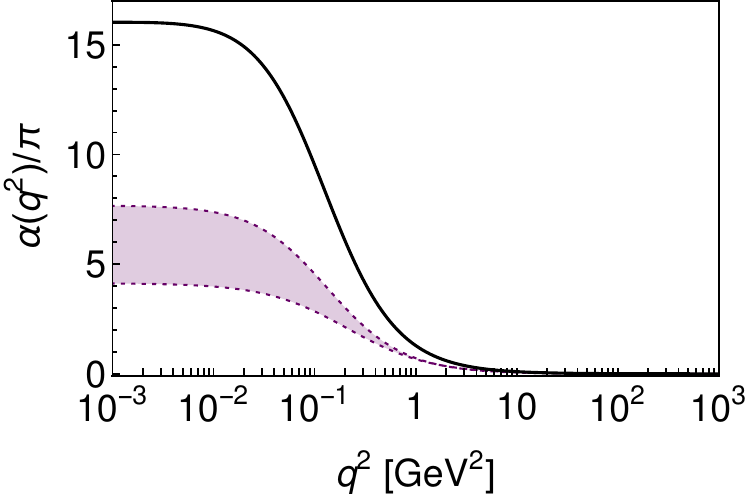}
        \caption{\textbf{[QC model]} Effective coupling parameterized as in Eq. \eqref{eq:coupLat}. The black line corresponds to the effective coupling associated with the bare vertex. Results for the other vertices lie within the band whose height is considerably diminished.}\label{fig:coup1}
\end{figure}

Another interesting measure of DCSB is given by the pseudoscalar meson leptonic decay constant, $f_\pi$. The chiral limit value can be easily computed from~\cite{Roberts:1994dr}:
    \begin{equation}
    \label{eq:fpi1}
       f_\pi^2 = \frac{3}{4\pi^2}  \int  dp^2 \frac{ p^2 Z(p^2) M(p^2)}{[p^2+M^2(p^2)]^2}  \left[M(p^2) - \frac{p^2}{2}M'(p^2) \right]
    \end{equation}
    and from the improved Pagels-Stokar-Cornwall formula derived in~\cite{Roberts:1994hh}:
    \begin{eqnarray}
        \nonumber
        f_\pi^2 &=& \frac{3}{8\pi^2} \int dp^2\; p^2 B^2(p^2)\Big(\sigma_v^2 -2 [\sigma_s \sigma_s'+p^2 \sigma_v \sigma_v'] \\
        \label{eq:fpi2}
        &-& p^2[\sigma_s \sigma_s'' - \sigma_s' \sigma_s']-p^4 [\sigma_v \sigma_v'' - \sigma_v' \sigma_v'] \Big)\;,
    \end{eqnarray}
where the dependence of $\sigma_{s,v}$ on $p^2$ has been omitted for notational convenience. We denote Eqs. \eqref{eq:fpi1}-\eqref{eq:fpi2} as F.1 and F.2, respectively. The obtained values are shown in Table~\ref{tab:fpi}. Unsurprisingly, for all dressed vertices employed herein, F.2 provides a better estimate of the chiral limit value of $f_\pi$ ($\approx 0.09$ GeV). For the BB vertex, both formulas produce very similar values (with a relative difference of $\sim 2-4\;\%$), while those obtained with BC and KP differ up to $12-16\;\%$. In general, as can be inferred from Tables~\ref{tab:m0} to \ref{tab:fpi}, BB vertex results exhibit less sensitivity to the choice of MT or QC interaction models. If the bare vertex is employed instead, F.1 gives  better estimates than the \emph{more accurate} F.2, which overestimates $f_\pi$ by $12-14\; \%$. In analogy to the relative largeness of $M(0)$ of the light quarks (of the bare vertex results with respect to the others), this feature could arise from the fact that the rainbow approximation requires larger infrared enhancement from the gluon model. To address this fact, we rewrite QC interaction as follows:
\begin{eqnarray}
\frac{\alpha_s(q^2)}{q^2} &\to& \frac{\tilde{\alpha}_s(q^2)}{q^2+m_g^2(q^2)}\;,\;m_g^2(q^2) = \frac{m_0^4}{q^2+m_0^2}\;,
\label{eq:coupLat}
\end{eqnarray}
where $\tilde{\alpha}_s(q^2)$ is parameterized as suggested in the combined SDE and lattice study~\cite{Aguilar:2010gm} such that $m_g(q^2)$ acts as a running-mass-like term and provides us with a \emph{gluon mass
scale}. Figure~\ref{fig:coup1} clearly shows that a considerable enhancement of the coupling $\alpha(0)/\pi$ is needed for the bare vertex. It is $3.5-5.5$ times larger than the corresponding values of the coupling for the other vertices to produce observed phenomenology. Note that it is despite the fact that $m_g(0)$ lies within a typical range $\approx 0.4 - 0.6$ GeV in all cases.

\begin{table}
\begin{ruledtabular}
\begin{tabular}{lcccc}
 Vertex  & F.1 & F.2 & F.1 & F.2 \\
\hline \\
& \multicolumn{2}{c}{MT Model~\cite{Maris:1999nt}}  & \multicolumn{2}{c}{QC Model~\cite{Qin:2011dd}} \\
\hline 
Bare & 0.088 & 0.101 & 0.088 & 0.106 \\
BC~\cite{Ball:1980ay} & 0.083 & 0.093 & 0.079 & 0.093\\
CP~\cite{Curtis:1990zs} & 0.082 & 0.091 & 0.082 & 0.094\\
KP~\cite{Kizilersu:2009kg} & 0.080 & 0.090 & 0.078 & 0.090\\
BB~\cite{Bashir:2011dp} & 0.085 & 0.100 & 0.088 & 0.103\\
\end{tabular}
\end{ruledtabular}
\caption{\label{tab:fpi} Chiral limit decay constants computed from Eqs. \eqref{eq:fpi1}-\eqref{eq:fpi2} (F.1 and F.2, respectively). Dimensioned quantities are expressed in GeV.}
\end{table}

\begin{figure}[ht!]
\centering
\includegraphics[width=0.45\textwidth]{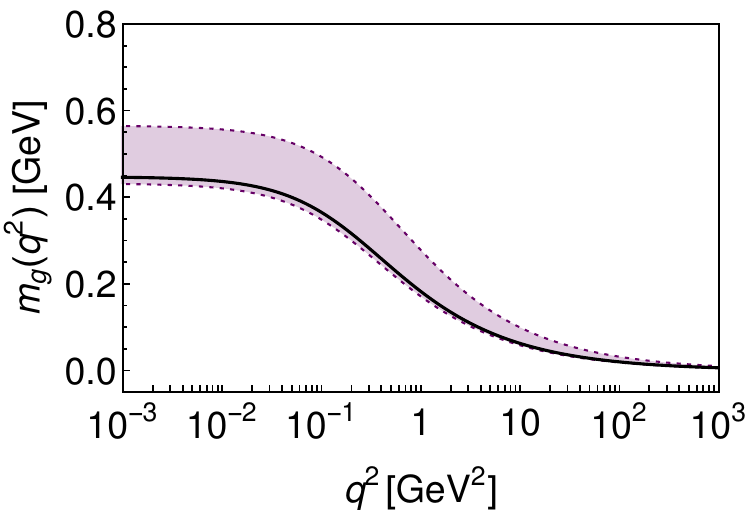}
        \caption{\textbf{[QC model]} The gluon \emph{running mass} from Eq. \eqref{eq:coupLat}. The black line corresponds to the effective coupling associated with the bare vertex. Results for the other vertices lie within the band.}\label{fig:coup2}
\end{figure}
Finally, motivated by the GellMann-Oakes-Renner relationship (see \cite{Chang:2011mu}, for example) and our values of chiral condensate and decay constants, one could argue that $m_\pi$ can be accurately obtained from realistic solutions of the Bethe-Salpeter equation, with a fully-consistent symmetry-preserving kernel. This is an outstanding challenge that we shall address elsewhere.

\section{Conclusions and Scope}
\label{sec:conclusions}

We have investigated the features of the dressed QGV and their impact on DCSB through the SDE for the quark propagator. Within a small phenomenologically sensible variation of the MT and QC model parameters, fixed solely by the chiral quark condensate, the results obtained from the refined vertex \emph{Ans\"atze} exhibit very similar quantitative behavior.
The robustness of the momentum-dependent mass function and the pion decay constant suggests that hadron observables could be accurately reproduced. Though the bare vertex results for the condensate and the decay constant compare well with other truncations, a notorious infrared enhancement (in the gluon models) is required. Firstly, recall that half of the structures which define the QGV can only contribute if chiral symmetry is dynamically broken. Secondly, there is a natural interplay between the role of the gluon propagator and the QGV. In order to generate required amount of DCSB, the bare vertex result depends on large infrared enhancement of the gluon propagator as it receives no such contribution from the dynamically generated vertex structures which are left out in this truncation scheme. 
A realistic and currently converging understanding of the gluon propagator can generate an acceptable running quark mass, via QCD’s gap equation, only as long as the QGV exhibits material infrared enhancement itself. Thus an intimate connection between the QGV and the DCSB is established. 
In this article, employing the MT and QC gluon models, we solve the gap equation using the following vertex \emph{Ans\"atze}: bare, BC, CP, KP and BB. All truncations described herein point towards the same qualitative pattern of DCSB. Expectedly, apart from the bare vertex, the infrared enhancement band of the mass function for all the other \emph{Ans\"atze} is rather narrow. Its width is what we expect to introduce error bars when we predict hadron observables using this formalism.   
 
The light quarks, weakly coupled to the Higgs field, owe their mass primarily to the infrared QCD dynamics. As one moves towards the heavy sector, weak mass generation commensurates with that coming from QCD's strong interactions; it is between the strange and charm quark masses (but closer to the former) that emergent and explicit mass generation have equal strength. These are qualitatively robust features of the SDE studies~\cite{Ding:2018xwy,Serna:2018dwk}, independent of the details of the truncation. 

To enhance the connection with hadron physics, it would be worth investigating if the vertex \emph{Ans\"atze} studied in this work are suitable for use in the non-perturbative studies of sophisticated hadron physics phenomenology in its fine details, the electromagnetic and transition form factors 

An immediate task would be writing a consistent Bethe-Salpeter kernel for all those vertices. It is known that, along with the bare QGV, a ladder-like kernel is sufficient for many needs, providing an accurate description of light pseudo-scalars and vector mesons (see for example,~\cite{Chang:2013pq,Raya:2015gva,Raya:2016yuj,Xu:2019ilh}). Nevertheless, for a fully-dressed QGV, the construction of a consistent Bethe-Salpeter kernel could be the next challenge~\cite{Chang:2009zb,Binosi:2016rxz}.  Moreover, DCSB generates a momentum-dependent dressed-quark anomalous chromomagnetic moment, which is large at infrared momenta and has an impact on the mass splitting between parity partners~\cite{Chang:2010hb,Bashir:2011dp,Lu:2017cln}. Thus, we strongly believe that the truncations which go beyond RL should be relevant for a variety of hadron properties, including the spectrum of the excited states, and the nucleon electromagnetic elastic and transition form factors such as~\cite{Eichmann:2016yit,Eichmann:2016hgl,Segovia:2015hra,Raya:2018ioy}. Some of those aspects are currently being investigated and will be reported elsewhere.


\section{Acknowledgements}
This research was partly supported by
{\em Coordinación de la Investigación Científica} (CIC) of the
University of Michoacan and CONACyT-Mexico 
through Grants No. 4.10 and CB2014-22117, respectively. KR acknowledges support from 
CONACyT-Mexico. FA acknowledges the financial support of HEC of Pakistan through Project No. 20-4500/NRPU/R$\&$D/HEC/14/727.

\bibliographystyle{unsrt}
\bibliography{bibliography}

\begin{thebibliography}{100}

\bibitem{Schwinger:1951ex}
Julian~S. Schwinger.
\newblock {On the Green's functions of quantized fields. 1.}
\newblock {\em Proc. Nat. Acad. Sci.}, 37:452--455, 1951.

\bibitem{Schwinger:1951hq}
Julian~S. Schwinger.
\newblock {On the Green's functions of quantized fields. 2.}
\newblock {\em Proc. Nat. Acad. Sci.}, 37:455--459, 1951.

\bibitem{Dyson:1949ha}
F.~J. Dyson.
\newblock {The S matrix in quantum electrodynamics}.
\newblock {\em Phys. Rev.}, 75:1736--1755, 1949.

\bibitem{Ball:1980ay}
James~S. Ball and Ting-Wai Chiu.
\newblock {Analytic Properties of the Vertex Function in Gauge Theories. 1.}
\newblock {\em Phys. Rev.}, D22:2542, 1980.

\bibitem{Curtis:1990zs}
D.~C. Curtis and M.~R. Pennington.
\newblock {Truncating the Schwinger-Dyson equations: How multiplicative
  renormalizability and the Ward identity restrict the three point vertex in
  QED}.
\newblock {\em Phys. Rev.}, D42:4165--4169, 1990.

\bibitem{Bashir:1994az}
A.~Bashir and M.~R. Pennington.
\newblock {Gauge independent chiral symmetry breaking in quenched QED}.
\newblock {\em Phys. Rev.}, D50:7679--7689, 1994.

\bibitem{Bashir:1995qr}
A.~Bashir and M.~R. Pennington.
\newblock {Constraint on the QED vertex from the mass anomalous dimension
  gamma(m) = 1}.
\newblock {\em Phys. Rev.}, D53:4694--4697, 1996.

\bibitem{Bashir:2011dp}
A.~Bashir, R.~Bermudez, L.~Chang, and C.D. Roberts.
\newblock {Dynamical chiral symmetry breaking and the fermion--gauge-boson
  vertex}.
\newblock {\em Phys.Rev.}, C85:045205, 2012.

\bibitem{Albino:2018ncl}
L.~Albino, A.~Bashir, L.~X.~Gutiérrez Guerrero, B.~El Bennich, and E.~Rojas.
\newblock {Transverse Takahashi Identities and Their Implications for Gauge
  Independent Dynamical Chiral Symmetry Breaking}.
\newblock {\em Phys. Rev.}, D100(5):054028, 2019.

\bibitem{Alkofer:2008et}
Reinhard Alkofer, Christian~S. Fischer, and Richard Williams.
\newblock {U(A)(1) anomaly and eta-prime mass from an infrared singular
  quark-gluon vertex}.
\newblock {\em Eur. Phys. J. A}, 38:53--60, 2008.

\bibitem{Kizilersu:2009kg}
A.~Kizilersu and M.R. Pennington.
\newblock {Building the Full Fermion-Photon Vertex of QED by Imposing
  Multiplicative Renormalizability of the Schwinger-Dyson Equations for the
  Fermion and Photon Propagators}.
\newblock {\em Phys.Rev.}, D79:125020, 2009.

\bibitem{Rojas:2013tza}
E.~Rojas, J.P.B.C. de~Melo, B.~El-Bennich, O.~Oliveira, and T.~Frederico.
\newblock {On the Quark-Gluon Vertex and Quark-Ghost Kernel: combining Lattice
  Simulations with Dyson-Schwinger equations}.
\newblock {\em JHEP}, 10:193, 2013.

\bibitem{Aguilar:2014lha}
A.~C. Aguilar, D.~Binosi, D.~Ibañez, and J.~Papavassiliou.
\newblock {New method for determining the quark-gluon vertex}.
\newblock {\em Phys. Rev.}, D90(6):065027, 2014.

\bibitem{Williams:2014iea}
Richard Williams.
\newblock {The quark-gluon vertex in Landau gauge bound-state studies}.
\newblock {\em Eur. Phys. J. A}, 51(5):57, 2015.

\bibitem{Gomez-Rocha:2015qga}
M.~Gomez-Rocha, T.~Hilger, and A.~Krassnigg.
\newblock {Effects of a dressed quark-gluon vertex in pseudoscalar heavy-light
  mesons}.
\newblock {\em Phys. Rev. D}, 92(5):054030, 2015.

\bibitem{Gomez-Rocha:2016cji}
M.~G\'omez-Rocha, T.~Hilger, and A.~Krassnigg.
\newblock {Effects of a dressed quark-gluon vertex in vector heavy-light mesons
  and theory average of the $B_c^*$ meson mass}.
\newblock {\em Phys. Rev. D}, 93(7):074010, 2016.

\bibitem{Binosi:2016wcx}
Daniele Binosi, Lei Chang, Joannis Papavassiliou, Si-Xue Qin, and Craig~D.
  Roberts.
\newblock {Natural constraints on the gluon-quark vertex}.
\newblock {\em Phys. Rev.}, D95(3):031501, 2017.

\bibitem{Bermudez:2017bpx}
R.~Bermudez, L.~Albino, L.~X. Guti\'errez-Guerrero, M.~E. Tejeda-Yeomans, and
  A.~Bashir.
\newblock {Quark-gluon Vertex: A Perturbation Theory Primer and Beyond}.
\newblock {\em Phys. Rev.}, D95(3):034041, 2017.

\bibitem{Aguilar:2018epe}
A.~C. Aguilar, J.~C. Cardona, M.~N. Ferreira, and J.~Papavassiliou.
\newblock {Quark gap equation with non-abelian Ball-Chiu vertex}.
\newblock {\em Phys. Rev.}, D98(1):014002, 2018.

\bibitem{Fischer:2006ub}
Christian~S. Fischer.
\newblock {Infrared properties of QCD from Dyson-Schwinger equations}.
\newblock {\em J. Phys. G}, 32:R253--R291, 2006.

\bibitem{Alkofer:2008tt}
Reinhard Alkofer, Christian~S. Fischer, Felipe~J. Llanes-Estrada, and Kai
  Schwenzer.
\newblock {The Quark-gluon vertex in Landau gauge QCD: Its role in dynamical
  chiral symmetry breaking and quark confinement}.
\newblock {\em Annals Phys.}, 324:106--172, 2009.

\bibitem{Bashir:2012fs}
Adnan Bashir, Lei Chang, Ian~C. Cloet, Bruno El-Bennich, Yu-Xin Liu, Craig~D.
  Roberts, and Peter~C. Tandy.
\newblock {Collective perspective on advances in Dyson-Schwinger Equation QCD}.
\newblock {\em Commun. Theor. Phys.}, 58:79--134, 2012.

\bibitem{Aznauryan:2012ba}
I.G. Aznauryan et~al.
\newblock {Studies of Nucleon Resonance Structure in Exclusive Meson
  Electroproduction}.
\newblock {\em Int. J. Mod. Phys. E}, 22:1330015, 2013.

\bibitem{Cloet:2013jya}
Ian~C. Cloet and Craig~D. Roberts.
\newblock {Explanation and Prediction of Observables using Continuum Strong
  QCD}.
\newblock {\em Prog. Part. Nucl. Phys.}, 77:1--69, 2014.

\bibitem{Horn:2016rip}
Tanja Horn and Craig~D. Roberts.
\newblock {The pion: an enigma within the Standard Model}.
\newblock {\em J. Phys.}, G43(7):073001, 2016.

\bibitem{Roberts:2020hiw}
Craig~D Roberts.
\newblock {Empirical Consequences of Emergent Mass}.
\newblock {\em Symmetry}, 12(9):1468, 2020.

\bibitem{Chang:2009zb}
Lei Chang and Craig~D. Roberts.
\newblock {Sketching the Bethe-Salpeter kernel}.
\newblock {\em Phys. Rev. Lett.}, 103:081601, 2009.

\bibitem{Qin:2011dd}
Si-xue Qin, Lei Chang, Yu-xin Liu, Craig~D. Roberts, and David~J. Wilson.
\newblock {Interaction model for the gap equation}.
\newblock {\em Phys. Rev.}, C84:042202, 2011.

\bibitem{Binosi:2016rxz}
Daniele Binosi, Lei Chang, Joannis Papavassiliou, Si-Xue Qin, and Craig~D.
  Roberts.
\newblock {Symmetry preserving truncations of the gap and Bethe-Salpeter
  equations}.
\newblock {\em Phys. Rev.}, D93(9):096010, 2016.

\bibitem{Slavnov:1972fg}
A.~A. Slavnov.
\newblock {Ward Identities in Gauge Theories}.
\newblock {\em Theor. Math. Phys.}, 10:99--107, 1972.
\newblock [Teor. Mat. Fiz.10,153(1972)].

\bibitem{Taylor:1971ff}
J.~C. Taylor.
\newblock {Ward Identities and Charge Renormalization of the Yang-Mills Field}.
\newblock {\em Nucl. Phys.}, B33:436--444, 1971.

\bibitem{Aslam:2015nia}
M.~Jamil Aslam, A.~Bashir, and L.~X. Gutierrez-Guerrero.
\newblock {Local Gauge Transformation for the Quark Propagator in an SU(N)
  Gauge Theory}.
\newblock {\em Phys. Rev.}, D93(7):076001, 2016.

\bibitem{DeMeerleer:2018txc}
T.~De~Meerleer, D.~Dudal, S.~P. Sorella, P.~Dall'Olio, and A.~Bashir.
\newblock {Fresh look at the Abelian and non-Abelian Landau-Khalatnikov-Fradkin
  transformations}.
\newblock {\em Phys. Rev.}, D97(7):074017, 2018.

\bibitem{DeMeerleer:2019kmh}
Tim De~Meerleer, David Dudal, Silvio~Paolo Sorella, Pietro Dall'Olio, and Adnan
  Bashir.
\newblock {Landau-Khalatnikov-Fradkin Transformations, Nielsen Identities,
  Their Equivalence and Implications for QCD}.
\newblock {\em Phys. Rev. D}, 101(8):085005, 2020.

\bibitem{Takahashi:1985yz}
Yasushi Takahashi.
\newblock {Canonical quantization and generalized ward relations: foundation of
  nonperturbative approach}.
\newblock In {\em {Positano Symp.1985:0019}}, page 0019, 1985.

\bibitem{Kondo:1996xn}
Kei-Ichi Kondo.
\newblock {Transverse Ward-Takahashi identity, anomaly and Schwinger-Dyson
  equation}.
\newblock {\em Int. J. Mod. Phys.}, A12:5651--5686, 1997.

\bibitem{He:2000we}
Han-Xin He, F.~C. Khanna, and Y.~Takahashi.
\newblock {Transverse Ward-Takahashi identity for the fermion boson vertex in
  gauge theories}.
\newblock {\em Phys. Lett.}, B480:222--228, 2000.

\bibitem{Qin:2013mta}
Si-Xue Qin, Lei Chang, Yu-Xin Liu, Craig~D. Roberts, and Sebastian~M. Schmidt.
\newblock {Practical corollaries of transverse Ward-Green-Takahashi
  identities}.
\newblock {\em Phys. Lett.}, B722:384--388, 2013.

\bibitem{Li:2019xwk}
Yi-Da Li and Qing Wang.
\newblock {Beyond Symmetries : Anomalies in Transverse Ward--Takahashi
  Identities}.
\newblock {\em Phys. Rev. D}, 102(5):056008, 2020.

\bibitem{Luo:2019ywn}
Cui-Bai Luo and Hong-Shi Zong.
\newblock {Transverse Ward-Takahashi identities and full vertex functions in
  different representations of QED$_3$}.
\newblock {\em Chin. Phys. C}, 44(7):073105, 2020.

\bibitem{Davydychev:2000rt}
Andrei~I. Davydychev, P.~Osland, and L.~Saks.
\newblock {Quark gluon vertex in arbitrary gauge and dimension}.
\newblock {\em Phys. Rev.}, D63:014022, 2001.

\bibitem{Bashir:1999bd}
A.~Bashir, A.~Kizilersu, and M.~R. Pennington.
\newblock {Analytic form of the one loop vertex and of the two loop fermion
  propagator in three-dimensional massless QED}.
\newblock 1999.

\bibitem{Bashir:2000rv}
A.~Bashir, A.~Kizilersu, and M.~R. Pennington.
\newblock {Does the weak coupling limit of the Burden-Tjiang deconstruction of
  the massless quenched three-dimensional QED vertex agree with perturbation
  theory?}
\newblock {\em Phys. Rev.}, D62:085002, 2000.

\bibitem{Bashir:2007qq}
A.~Bashir, Y.~Concha-Sanchez, and Robert Delbourgo.
\newblock {3-point off-shell vertex in scalar QED in arbitrary gauge and
  dimension}.
\newblock {\em Phys. Rev. D}, 76:065009, 2007.

\bibitem{Bashir:2011vg}
Adnan Bashir, Alfredo Raya, and Saul Sanchez-Madrigal.
\newblock {Chiral Symmetry Breaking and Confinement Beyond Rainbow-Ladder
  Truncation}.
\newblock {\em Phys. Rev.}, D84:036013, 2011.

\bibitem{Chang:2011vu}
Lei Chang, Craig~D. Roberts, and Peter~C. Tandy.
\newblock {Selected highlights from the study of mesons}.
\newblock {\em Chin. J. Phys.}, 49:955--1004, 2011.

\bibitem{Eichmann:2016yit}
Gernot Eichmann, Helios Sanchis-Alepuz, Richard Williams, Reinhard Alkofer, and
  Christian~S. Fischer.
\newblock {Baryons as relativistic three-quark bound states}.
\newblock {\em Prog. Part. Nucl. Phys.}, 91:1--100, 2016.

\bibitem{Eichmann:2016hgl}
Gernot Eichmann, Christian~S. Fischer, and Helios Sanchis-Alepuz.
\newblock {Light baryons and their excitations}.
\newblock {\em Phys. Rev.}, D94(9):094033, 2016.

\bibitem{Qin:2014vya}
Si-Xue Qin, Craig~D. Roberts, and Sebastian~M. Schmidt.
\newblock {Ward–Green–Takahashi identities and the axial-vector vertex}.
\newblock {\em Phys. Lett.}, B733:202--208, 2014.

\bibitem{Maris:1999nt}
Pieter Maris and Peter~C. Tandy.
\newblock {Bethe-Salpeter study of vector meson masses and decay constants}.
\newblock {\em Phys. Rev.}, C60:055214, 1999.

\bibitem{Aguilar:2004sw}
A.~C. Aguilar and A.~A. Natale.
\newblock {A Dynamical gluon mass solution in a coupled system of the
  Schwinger-Dyson equations}.
\newblock {\em JHEP}, 08:057, 2004.

\bibitem{Cucchieri:2007md}
Attilio Cucchieri and Tereza Mendes.
\newblock {What's up with IR gluon and ghost propagators in Landau gauge? A
  puzzling answer from huge lattices}.
\newblock {\em PoS}, LAT2007:297, 2007.

\bibitem{Bogolubsky:2007ud}
I.~L. Bogolubsky, E.~M. Ilgenfritz, M.~Muller-Preussker, and A.~Sternbeck.
\newblock {The Landau gauge gluon and ghost propagators in 4D SU(3)
  gluodynamics in large lattice volumes}.
\newblock {\em PoS}, LAT2007:290, 2007.

\bibitem{Cucchieri:2010xr}
Attilio Cucchieri and Tereza Mendes.
\newblock {Numerical test of the Gribov-Zwanziger scenario in Landau gauge}.
\newblock {\em PoS}, QCD-TNT09:026, 2009.

\bibitem{Bogolubsky:2009dc}
I.~L. Bogolubsky, E.~M. Ilgenfritz, M.~Muller-Preussker, and A.~Sternbeck.
\newblock {Lattice gluodynamics computation of Landau gauge Green's functions
  in the deep infrared}.
\newblock {\em Phys. Lett.}, B676:69--73, 2009.

\bibitem{Aguilar:2008xm}
A.C. Aguilar, D.~Binosi, and J.~Papavassiliou.
\newblock {Gluon and ghost propagators in the Landau gauge: Deriving lattice
  results from Schwinger-Dyson equations}.
\newblock {\em Phys.Rev.}, D78:025010, 2008.

\bibitem{Boucaud:2008ky}
Philippe Boucaud, J.P. Leroy, A.~Le~Yaouanc, J.~Micheli, O.~Pene, et~al.
\newblock {On the IR behaviour of the Landau-gauge ghost propagator}.
\newblock {\em JHEP}, 0806:099, 2008.

\bibitem{Fischer:2008uz}
Christian~S. Fischer, Axel Maas, and Jan~M. Pawlowski.
\newblock {On the infrared behavior of Landau gauge Yang-Mills theory}.
\newblock {\em Annals Phys.}, 324:2408--2437, 2009.

\bibitem{Aguilar:2009nf}
A.~C. Aguilar, D.~Binosi, J.~Papavassiliou, and J.~Rodriguez-Quintero.
\newblock {Non-perturbative comparison of QCD effective charges}.
\newblock {\em Phys. Rev.}, D80:085018, 2009.

\bibitem{Pennington:2011xs}
M.R. Pennington and D.J. Wilson.
\newblock {Are the Dressed Gluon and Ghost Propagators in the Landau Gauge
  presently determined in the confinement regime of QCD?}
\newblock {\em Phys.Rev.}, D84:119901, 2011.

\bibitem{Blum:2014gna}
Adrian Blum, Markus~Q. Huber, Mario Mitter, and Lorenz von Smekal.
\newblock {Gluonic three-point correlations in pure Landau gauge QCD}.
\newblock {\em Phys. Rev.}, D89:061703, 2014.

\bibitem{Cyrol:2016tym}
Anton~K. Cyrol, Leonard Fister, Mario Mitter, Jan~M. Pawlowski, and Nils
  Strodthoff.
\newblock {Landau gauge Yang-Mills correlation functions}.
\newblock {\em Phys. Rev.}, D94(5):054005, 2016.

\bibitem{Huber:2017txg}
Markus~Q. Huber.
\newblock {On non-primitively divergent vertices of Yang–Mills theory}.
\newblock {\em Eur. Phys. J.}, C77(11):733, 2017.

\bibitem{Dudal:2007cw}
D.~Dudal, S.~P. Sorella, N.~Vandersickel, and H.~Verschelde.
\newblock {New features of the gluon and ghost propagator in the infrared
  region from the Gribov-Zwanziger approach}.
\newblock {\em Phys. Rev.}, D77:071501, 2008.

\bibitem{Dudal:2008sp}
David Dudal, John~A. Gracey, Silvio~Paolo Sorella, Nele Vandersickel, and Henri
  Verschelde.
\newblock {A Refinement of the Gribov-Zwanziger approach in the Landau gauge:
  Infrared propagators in harmony with the lattice results}.
\newblock {\em Phys. Rev.}, D78:065047, 2008.

\bibitem{Dudal:2010tf}
D.~Dudal, O.~Oliveira, and N.~Vandersickel.
\newblock {Indirect lattice evidence for the Refined Gribov-Zwanziger formalism
  and the gluon condensate $\langle{A^2}\rangle$ in the Landau gauge}.
\newblock {\em Phys. Rev.}, D81:074505, 2010.

\bibitem{Cornwall:1981zr}
John~M. Cornwall.
\newblock {Dynamical Mass Generation in Continuum QCD}.
\newblock {\em Phys. Rev.}, D26:1453, 1982.

\bibitem{Bowman:2007du}
Patrick~O. Bowman, Urs~M. Heller, Derek~B. Leinweber, Maria~B. Parappilly,
  Andre Sternbeck, Lorenz von Smekal, Anthony~G. Williams, and Jian-bo Zhang.
\newblock {Scaling behavior and positivity violation of the gluon propagator in
  full QCD}.
\newblock {\em Phys. Rev.}, D76:094505, 2007.

\bibitem{Ayala:2012pb}
A.~Ayala, A.~Bashir, D.~Binosi, M.~Cristoforetti, and J.~Rodriguez-Quintero.
\newblock {Quark flavour effects on gluon and ghost propagators}.
\newblock {\em Phys.Rev.}, D86:074512, 2012.

\bibitem{Aguilar:2012rz}
A.~C. Aguilar, D.~Binosi, and J.~Papavassiliou.
\newblock {Unquenching the gluon propagator with Schwinger-Dyson equations}.
\newblock {\em Phys. Rev.}, D86:014032, 2012.

\bibitem{Bashir:2013zha}
A.~Bashir, A.~Raya, and J.~Rodriguez-Quintero.
\newblock {QCD: Restoration of Chiral Symmetry and Deconfinement for Large
  $N_f$}.
\newblock {\em Phys.Rev.}, D88:054003, 2013.

\bibitem{Binosi:2016nme}
Daniele Binosi, Cedric Mezrag, Joannis Papavassiliou, Craig~D. Roberts, and
  Jose Rodriguez-Quintero.
\newblock {Process-independent strong running coupling}.
\newblock {\em Phys. Rev.}, D96(5):054026, 2017.

\bibitem{Cui:2019dwv}
Zhu-Fang Cui, Jin-Li Zhang, Daniele Binosi, Feliciano de~Soto, Cédric Mezrag,
  Joannis Papavassiliou, Craig~D Roberts, Jose Rodríguez-Quintero, Jorge
  Segovia, and Savvas Zafeiropoulos.
\newblock {Effective charge from lattice QCD}.
\newblock 12 2019.

\bibitem{Cucchieri:2009kk}
Attilio Cucchieri, Tereza Mendes, and Elton M.~S. Santos.
\newblock {Covariant gauge on the lattice: A New implementation}.
\newblock {\em Phys. Rev. Lett.}, 103:141602, 2009.

\bibitem{Boucaud:2018xup}
Ph~Boucaud, F.~De~Soto, K.~Raya, J.~Rodríguez-Quintero, and S.~Zafeiropoulos.
\newblock {Discretization effects on renormalized gauge-field Green's
  functions, scale setting and gluon mass}.
\newblock 2018.

\bibitem{Kern:2019nzx}
Wolfgang Kern, Markus~Q. Huber, and Reinhard Alkofer.
\newblock {Spectral dimension as a tool for analyzing nonperturbative
  propagators}.
\newblock {\em Phys. Rev. D}, 100(9):094037, 2019.

\bibitem{Chang:2013pq}
Lei Chang, I.~C. Cloet, J.~J. Cobos-Martinez, C.~D. Roberts, S.~M. Schmidt, and
  P.~C. Tandy.
\newblock {Imaging dynamical chiral symmetry breaking: pion wave function on
  the light front}.
\newblock {\em Phys. Rev. Lett.}, 110(13):132001, 2013.

\bibitem{Gao:2017uox}
Fei Gao, Si-Xue Qin, Craig~D. Roberts, and Jose Rodriguez-Quintero.
\newblock {Locating the Gribov horizon}.
\newblock {\em Phys. Rev.}, D97(3):034010, 2018.

\bibitem{Qin:2018dqp}
Si-Xue Qin, Craig~D. Roberts, and Sebastian~M. Schmidt.
\newblock {Poincar\'e-covariant analysis of heavy-quark baryons}.
\newblock {\em Phys. Rev.}, D97(11):114017, 2018.

\bibitem{Qin:2019hgk}
Si-xue Qin, Craig~D Roberts, and Sebastian~M Schmidt.
\newblock {Spectrum of light- and heavy-baryons}.
\newblock {\em Few Body Syst.}, 60(2):26, 2019.

\bibitem{Nguyen:2011jy}
Trang Nguyen, Adnan Bashir, Craig~D. Roberts, and Peter~C. Tandy.
\newblock {Pion and kaon valence-quark parton distribution functions}.
\newblock {\em Phys. Rev.}, C83:062201, 2011.

\bibitem{Ding:2019lwe}
Minghui Ding, Khépani Raya, Daniele Binosi, Lei Chang, Craig~D Roberts, and
  Sebastian~M. Schmidt.
\newblock {Symmetry, symmetry breaking, and pion parton distributions}.
\newblock {\em Phys. Rev. D}, 101(5):054014, 2020.

\bibitem{Chang:2013nia}
L.~Chang, I.~C. Cloët, C.~D. Roberts, S.~M. Schmidt, and P.~C. Tandy.
\newblock {Pion electromagnetic form factor at spacelike momenta}.
\newblock {\em Phys. Rev. Lett.}, 111(14):141802, 2013.

\bibitem{Raya:2015gva}
K.~Raya, L.~Chang, A.~Bashir, J.~J. Cobos-Martinez, L.~X. Gutiérrez-Guerrero,
  C.~D. Roberts, and P.~C. Tandy.
\newblock {Structure of the neutral pion and its electromagnetic transition
  form factor}.
\newblock {\em Phys. Rev.}, D93(7):074017, 2016.

\bibitem{Eichmann:2017wil}
Gernot Eichmann, Christian~S. Fischer, Esther Weil, and Richard Williams.
\newblock {On the large-$Q^2$ behavior of the pion transition form factor}.
\newblock {\em Phys. Lett. B}, 774:425--429, 2017.

\bibitem{Raya:2016yuj}
Khepani Raya, Minghui Ding, Adnan Bashir, Lei Chang, and Craig~D. Roberts.
\newblock {Partonic structure of neutral pseudoscalars via two photon
  transition form factors}.
\newblock {\em Phys. Rev.}, D95(7):074014, 2017.

\bibitem{Ding:2018xwy}
Minghui Ding, Khepani Raya, Adnan Bashir, Daniele Binosi, Lei Chang, Muyang
  Chen, and Craig~D. Roberts.
\newblock {$\gamma^\ast \gamma \to \eta, \eta^\prime$ transition form factors}.
\newblock {\em Phys. Rev.}, D99(1):014014, 2019.

\bibitem{Raya:2019dnh}
Khépani Raya, Adnan Bashir, and Pablo Roig.
\newblock {Contribution of neutral pseudoscalar mesons to $a_\mu^{HLbL}$ within
  a Schwinger-Dyson equations approach to QCD}.
\newblock {\em Phys. Rev. D}, 101(7):074021, 2020.

\bibitem{Eichmann:2019bqf}
Gernot Eichmann, Christian~S. Fischer, and Richard Williams.
\newblock {Kaon-box contribution to the anomalous magnetic moment of the muon}.
\newblock {\em Phys. Rev. D}, 101(5):054015, 2020.

\bibitem{Eichmann:2019tjk}
Gernot Eichmann, Christian~S. Fischer, Esther Weil, and Richard Williams.
\newblock {Single pseudoscalar meson pole and pion box contributions to the
  anomalous magnetic moment of the muon}.
\newblock {\em Phys. Lett. B}, 797:134855, 2019.
\newblock [Erratum: Phys.Lett.B 799, 135029 (2019)].

\bibitem{Aoyama:2020ynm}
T.~Aoyama et~al.
\newblock {The anomalous magnetic moment of the muon in the Standard Model}.
\newblock {\em Phys. Rept.}, 887:1--166, 2020.

\bibitem{Ward:1950xp}
John~Clive Ward.
\newblock {An Identity in Quantum Electrodynamics}.
\newblock {\em Phys. Rev.}, 78:182, 1950.

\bibitem{Green:1953te}
H.S. Green.
\newblock {A Pre-renormalized quantum electrodynamics}.
\newblock {\em Proc. Phys. Soc. A}, 66:873--880, 1953.

\bibitem{Fradkin:1955jr}
E.S. Fradkin.
\newblock {Concerning some general relations of quantum electrodynamics}.
\newblock {\em Zh. Eksp. Teor. Fiz.}, 29:258--261, 1955.

\bibitem{Takahashi:1957xn}
Y.~Takahashi.
\newblock {On the generalized Ward identity}.
\newblock {\em Nuovo Cim.}, 6:371, 1957.

\bibitem{Chang:2010hb}
Lei Chang, Yu-Xin Liu, and Craig~D. Roberts.
\newblock {Dressed-quark anomalous magnetic moments}.
\newblock {\em Phys. Rev. Lett.}, 106:072001, 2011.

\bibitem{Chang:2011ei}
Lei Chang and Craig~D. Roberts.
\newblock {Tracing masses of ground-state light-quark mesons}.
\newblock {\em Phys. Rev.}, C85:052201, 2012.

\bibitem{Kizilersu:1995iz}
A.~Kizilersu, M.~Reenders, and M.~R. Pennington.
\newblock {One loop QED vertex in any covariant gauge: Its complete analytic
  form}.
\newblock {\em Phys. Rev.}, D52:1242--1259, 1995.

\bibitem{Kizilersu:2013hea}
Ayse Kizilersu, Tom Sizer, and Anthony~G. Williams.
\newblock {Strongly-Coupled Unquenched QED4 Propagators Using Schwinger-Dyson
  Equations}.
\newblock {\em Phys. Rev.}, D88:045008, 2013.

\bibitem{Williams:2006vva}
R.~Williams, C.~S. Fischer, and M.~R. Pennington.
\newblock {Anti-q q condensate for light quarks beyond the chiral limit}.
\newblock {\em Phys. Lett.}, B645:167--172, 2007.

\bibitem{Brodsky:2010xf}
Stanley~J. Brodsky, Craig~D. Roberts, Robert Shrock, and Peter~C. Tandy.
\newblock {Essence of the vacuum quark condensate}.
\newblock {\em Phys. Rev.}, C82:022201, 2010.

\bibitem{Chang:2011mu}
Lei Chang, Craig~D. Roberts, and Peter~C. Tandy.
\newblock {Expanding the concept of in-hadron condensates}.
\newblock {\em Phys. Rev.}, C85:012201, 2012.

\bibitem{Roberts:1994dr}
Craig~D. Roberts and Anthony~G. Williams.
\newblock {Dyson-Schwinger equations and their application to hadronic
  physics}.
\newblock {\em Prog. Part. Nucl. Phys.}, 33:477--575, 1994.

\bibitem{Serna:2018dwk}
Fernando~E. Serna, Chen Chen, and Bruno El-Bennich.
\newblock {Interplay of dynamical and explicit chiral symmetry breaking effects
  on a quark}.
\newblock {\em Phys. Rev. D}, 99(9):094027, 2019.

\bibitem{Roberts:1994hh}
Craig~D. Roberts.
\newblock {Electromagnetic pion form-factor and neutral pion decay width}.
\newblock {\em Nucl. Phys.}, A605:475--495, 1996.

\bibitem{Aguilar:2010gm}
A.~C. Aguilar, D.~Binosi, and J.~Papavassiliou.
\newblock {QCD effective charges from lattice data}.
\newblock {\em JHEP}, 07:002, 2010.

\bibitem{Xu:2019ilh}
Yin-Zhen Xu, Daniele Binosi, Zhu-Fang Cui, Bo-Lin Li, Craig~D Roberts,
  Shu-Sheng Xu, and Hong~Shi Zong.
\newblock {Elastic electromagnetic form factors of vector mesons}.
\newblock {\em Phys. Rev. D}, 100(11):114038, 2019.

\bibitem{Lu:2017cln}
Ya~Lu, Chen Chen, Craig~D. Roberts, Jorge Segovia, Shu-Sheng Xu, and Hong-Shi
  Zong.
\newblock {Parity partners in the baryon resonance spectrum}.
\newblock {\em Phys. Rev.}, C96(1):015208, 2017.

\bibitem{Segovia:2015hra}
Jorge Segovia, Bruno El-Bennich, Eduardo Rojas, Ian~C. Cloet, Craig~D. Roberts,
  Shu-Sheng Xu, and Hong-Shi Zong.
\newblock {Completing the picture of the Roper resonance}.
\newblock {\em Phys. Rev. Lett.}, 115(17):171801, 2015.

\bibitem{Raya:2018ioy}
K.~Raya, L.~X. Guti\'errez, and A.~Bashir.
\newblock {Structure of the orbital excited $N^*$ from the Schwinger-Dyson
  equations}.
\newblock {\em Few Body Syst.}, 59(5):89, 2018.

\end{thebibliography}

\end{document}